\documentclass[oupdraft]{bio}
\usepackage{url}
\usepackage{amsmath}
\usepackage{amssymb}
\usepackage{latexsym}
\usepackage{graphics}
\usepackage{graphicx}
\usepackage{epsfig,epsf,rotating}
\usepackage{epsf}
\usepackage{theorem}
\usepackage{bm}
\usepackage{booktabs}
\usepackage{longtable}
\usepackage{subfig}
\usepackage{multirow}
\usepackage{array}
\newcommand{\I}{\text{I}}
\newcommand{\NB}{\text{NB}}
\newcommand{\IG}{\text{IG}}

\newcommand{\normal}{\text{N}}

\newcommand{\Ga}{\text{Ga}}

\newcommand\Tstrut{\rule{0pt}{3ex}} 
\usepackage[a4paper,top=1.8in,bottom = 1.8in,
left=1in,right=1in]{geometry}
\usepackage[usenames,dvipsnames]{color}
\usepackage{color,soul}
\usepackage{xcolor,framed}
\definecolor{shadecolor}{HTML}{FFF200}
\usepackage{ifthen}
\newboolean{isclean}
\setboolean{isclean}{false}
\ifthenelse{\boolean{isclean}}{

}{}
\graphicspath{{fig_main/}}
\usepackage[usenames,dvipsnames]{color}
\usepackage{color,soul}
\usepackage{xcolor,framed}
\definecolor{shadecolor}{HTML}{FFF200}

\newcommand{\ech}{\color{black}\rm }
\newcommand{\bch}{\color{black} }  
\renewcommand{\ech }{\color{black}\rm }  

\usepackage{natbib}

\history{Received August 1, 2010;
revised October 1, 2010;
accepted for publication November 1, 2010}

\begin{document}

\title{A Bayesian Zero-Inflated Negative Binomial Regression Model for the Integrative Analysis of Microbiome Data  }
\author{
	Shuang Jiang\\
	\vspace*{-.1in} 
\textit{Department of Statistical Science, Southern Methodist University, Dallas, TX 75275}
	\\[2pt]
	Guanghua Xiao\\
	\vspace*{.05in}
	\textit{Quantitative Biomedical Research Center, University of Texas Southwestern Medical Center, \\ \vspace*{-.12in} Dallas, TX 75390}
	\\[2pt]
	Andrew Y. Koh\\
	\vspace*{.05in}
	\textit{Departments of Pediatrics and Microbiology, University of Texas Southwestern Medical Center, \\ \vspace*{-.12in} Dallas, TX 75390}
	\\[2pt]
	Qiwei Li$^\ast$\\
		\vspace*{.05in}
\textit{Department of Mathematical Sciences, The University of Texas at Dallas, \\ \vspace*{-.12in} Richardson, TX 75080}\\
	{Qiwei.Li@UTDallas.edu} 
	\\[2pt]
	Xiaowei Zhan$^\ast$\\
	\vspace*{.05in}
\textit{Quantitative Biomedical Research Center, University of Texas Southwestern Medical Center,\\
	\vspace*{-.12in} Dallas, TX 75390}\\
	{Xiaowei.Zhan@UTSouthwestern.edu}
	\\[2pt]
}

\markboth%
{S. Jiang and others}
{A Bayesian Zero-Inflated Negative Binomial Integrative Model}

\maketitle

\footnotetext{To whom correspondence should be addressed.}

\begin{abstract}
{Microbiome `omics approaches can reveal intriguing relationships between the human microbiome and certain disease states. Along with identification of specific bacteria taxa associated with diseases, recent scientific advancements provide mounting evidence that metabolism, genetics and environmental factors can all modulate these microbial effects. However, the current methods for integrating microbiome data and other covariates are severely lacking.  Hence, we present an integrative Bayesian zero-inflated negative binomial regression model that can both distinguish differentially abundant taxa with distinct phenotypes and quantify covariate-taxa effects. Our model demonstrates good performance using simulated data. Furthermore, we successfully integrated microbiome taxonomies and metabolomics in two real microbiome datasets to provide biologically interpretable findings. In all, we proposed a novel integrative Bayesian regression model that features bacterial differential abundance analysis and microbiome-covariate effects quantifications, which makes it suitable for general microbiome studies. }
{Bayesian regression; Count data; Feature selection; Integrative analysis; Microbiome; Mixture models; Zero-inflated negative binomial model}

\end{abstract}

\section{Introduction}\label{introduction}
The human microbiome is estimated to contain $3.0\times10^{13}$ bacteria \citep{sender2016revised} and $3.3\times10^6$ microbial genes \citep{qin2010human}. Microbial communities have a profound impact on human health \citep{ursell2012defining}. Recently, microbiome studies have identified disease-associated bacteria taxa in type 2 diabetes \citep{karlsson2013gut}, liver cirrhosis \citep{qin2014}, inflammatory bowel disease \citep{halfvarson2017dynamics}, and melanoma patients responsive to cancer immunotherapy \citep{frankel2017}. An increasing number of research projects continue to systematically investigate the role of the microbiome in human diseases \citep{integrative2014integrative}.

While innovations in next-generation sequencing technology continue to shape the next steps in the microbiome field, the statistical methods used in microbiome research have not kept pace. For instance, metagenomic shotgun sequencing (MSS) generates a massive amount of sequence reads that can provide species or isolate level taxonomic resolution \citep{segata2012metagenomic}. The subsequent comparative
statistical analysis assesses whether specific species are associated with a phenotypic state or experimental condition.

Upon surveying commonly used statistical approaches, one method focuses on the comparison of multi-taxa \citep{chen2012associating,kelly2015power,zhao2015testing,wu2016adaptive}, frequently termed the microbiome community. However, those approaches do not aim to identify differentially abundant species---making clinical interpretation, mechanistic insights, and biological validations difficult. Another approach interrogates each individual bacteria taxa for different groups or conditions. For example, \cite{la2015hypothesis} utilizes a Wilcoxon rank-sum test or Kruskal-Wallis test for groupwise comparisons on microbiome compositional data. Recently, methods developed for RNA-seq data have been adapted to microbiome studies, e.g. the negative-binomial regression model in \texttt{DESeq2} \citep{love2014moderated} and overdispersed Poisson model in \texttt{edgeR} \citep{robinson2010edger}. These methods, however, are not optimized for microbiome datasets.

Microbial abundance can be affected by covariates, such as metabolites, antibiotics and host genetics. These confounding variables need to be adjusted for more accurate differential abundance analysis. Ultimately, there may be a clinical need to quantify the associations between microbiome and clinical confounders \citep{kinross2011gut,zhu2018precision,maier2018extensive}. One common approach is to calculate pairwise correlations between all taxa and covariates \citep{li2008symbiotic}, but this method may be significantly underpowered. Other model-based methods \citep{chen2013variable,wadsworth2017integrative} have been proposed to detect covariate-taxa associations, but the taxon-outcome associations have been ignored. \bch Recently, \cite{li2018conditional} developed a multivariate zero-inflated logistic-normal model to quantify the associations between microbiome abundances and multiple factors (e.g. disease risk factors or health outcomes) based on microbiome compositional data instead of the count data. \ech

Here, we propose a Bayesian integrative model to analyze microbiome count data. Our model jointly identifies differentially abundant taxa among multiple groups and simultaneously quantifies the taxon-covariate associations. Our modeling construction includes several advantages. First, it characterizes the over-dispersion and zero-inflation frequently observed in microbiome count data by introducing a zero-inflated negative binomial (ZINB) model. Second, it models the heterogeneity from different sequencing depths, covariate effects, and group effects via a log-linear regression framework on the ZINB mean components. Last, we propose two feature selection processes to simultaneously detect differentially abundant taxa and estimate the covariate-taxa associations using the \textit{spike-and-slab} priors. We compute Bayesian posterior probabilities for these correlated features and provide the Bayesian false discovery rate (FDR). Extensive and thorough use of simulated data demonstrates that our model largely improved performance when compared with existing methods. We present two applications of real microbiome datasets with various covariate sets. Biological interpretations of our results confirm those of previous studies and offer a more comprehensive understanding of the underlying mechanism in disease etiology.

The paper is organized as follows. In Section \ref{hierarchical}, we introduce the integrative hierarchical mixture model and the prior formulations. Section \ref{fitting} supplies a brief discussion of the MCMC algorithm and the resulting posterior inference. In Section \ref{simulation}, we evaluate model performance on simulated data through a comparison study. We investigate the covariate association in Section \ref{association-analysis}. Two real data analyses are shown in Section \ref{realdata}. Our conclusions are presented in Section \ref{conclusion}.
\section{Hierarchical Model}\label{hierarchical}
Our model starts with a high-dimensional count matrix where each entry represents the count of sequence reads belonging to a taxonomy such as bacterial species. Specifically, we denote $\bm{Y}_{n \times p}$ (usually $n\ll p$) be a microbial abundance matrix, with $y_{ij}\in\mathbb{N},i=1,\ldots,n,j=1,\ldots,p$ representing the observed count of the $i$-th sample and $j$-th taxon out of the total $n$ samples and $p$ taxa (features). Note that the proposed model can also be applied to an operational taxonomic unit (OTU) count table obtained via 16S metagenomic approaches. For an OTU table, each feature would be a taxonomic unit of a bacteria species or genus depending on the sequence similarity threshold (e.g. $97\%$).We also denote a covariate matrix $\bm{X}_{n\times R}$ where each entry $x_{ir}$ represents the measurement of the $r$-th covariate on the $i$-th sample. The graphical formulation of the proposed model is summarized in Figure S1 and S2 in the supplement.


\subsection{Count generating process}
In practice, the microbial abundance matrix $\bm{Y}$ is characterized by an inflated amount of zeros, resulting from insufficient sampling depth. Meanwhile, the abundance matrix usually consists of extremely large counts. Based on these two facts, we assume that each count is sampled from a zero-inflated negative binomial (ZINB) distribution so as to simultaneously account for both zero-inflation and over-dispersion presented in $\bm{Y}$:
\begin{equation}\label{y-eq}
	y_{ij}|\pi,\lambda_{ij},\phi_j\quad\sim\quad\pi \I(y_{ij}=0)+(1-\pi)\text{NB}(y_{ij};\lambda_{ij},\phi_j),
\end{equation}
where $\pi \in [0,1]$ represents the weight of generating extra zeros, $\I(\cdot)$ is an indicator function, and $\text{NB}(y;\lambda, \phi)$ denotes a negative binomial distribution for random variable $y$ with the expectation $\lambda$ and dispersion $1/\phi$. Under this parameterization, the variance of $y$ is $\lambda+\lambda^2/\phi$. A small value of $\phi$ allows modeling of extra variation. Note that increasing $\phi$ towards infinity yields a Poisson distribution with both expectation and variance equal to $\lambda$. We assume a Gamma prior $\text{Ga}(a_{\phi},b_{\phi})$ for the dispersion parameter $\phi$. 

An equivalent way to model this count generating process is to introduce a latent binary variable $r_{ij}$, such that
\begin{equation}\label{y-eq2}
	y_{ij}|r_{ij},\lambda_{ij},\phi_j\begin{cases}{\begin{array}{ll}\quad\sim\quad\text{NB}(\lambda_{ij},\phi_j) & \text{ if } r_{ij}=0\\\quad=\quad0 & \text{ if } r_{ij}=1\end{array}}\end{cases},
\end{equation}
where $r_{ij}$ is from a Bernoulli distribution with parameter $\pi$, i.e. $\sim\text{Bernoulli}(\pi)$. We further impose $\pi\sim\text{Beta}(a_{\pi}, b_{\pi})$, which leads to a Beta-Bernoulli prior for $r_{ij}$ with expectation $a_\pi/(a_\pi+b_\pi)$. 

\subsection{Integrative modeling with feature selection}
Microbiome count data is characterized by high variability in the number of reads among samples from different groups (due to distinct biological conditions), or even the same group (due to uneven sequencing depths). To accommodate this setting, we parameterize the mean parameter $\lambda_{ij}$ of the negative binomial distribution as the multiplicative effects of two positive random effects: 1) the size factor $s_i$ reflects how the sequencing depth affects counts across all taxa observed in the $i$-th sample; 2) the \bch normalized \ech abundance $\alpha_{ij}$ for the $j$-th taxon in the $i$-th sample once the sample-specific variability has been accounted for. Our goal is to find a subset of $p$ taxa that enables us to discriminate the $n$ samples from $K$ distinct groups. We introduce a binary latent vector $\bm{\gamma} =(\gamma_1,\ldots,\gamma_p)$, with $\gamma_j=1$ indicating that the $j$-th taxon has significantly differential abundances among the $K$ groups, and $\gamma_j=0$ otherwise. Therefore, conditional on $r_{ij}=0$, we reparameterize the \bch negative binomial \ech kernel of Equation (\ref{y-eq2}) as follows:
\begin{equation}\label{intergrate}
	y_{ij}|r_{ij} = 0, \gamma_j, s_i,\alpha_{ijk},\alpha_{ij0}\quad\sim\quad\begin{cases}\begin{array}{ll}
			\text{NB}(y_{ij};s_i\alpha_{ij0},\phi_j) & \text{if}~ \gamma_j  =0\\
			\text{NB}(y_{ij};s_i\alpha_{ijk},\phi_j) & \text{if}~ \gamma_j  =1 \text{ and } z_i=k
	\end{array}\end{cases}.
\end{equation}
Here, $z_i$ is the sample allocation indicator. Collectively, $\bm{z}_{n \times 1} = (z_1, z_2, \ldots, z_n)^T$ indicates the membership for each sample, where $z_i = k, k \in \{1, \ldots, K\}$ reveals that the $i$-th sample belongs to the $k$-th group. $s_i$ is the size factor of the $i$-th sample, which can be estimated from the data (see Section \ref{size factor}). We assume an independent Bernoulli prior $\gamma_j \sim\text{Bernoulli}(\omega)$ for each taxon $j$, and further impose a beta hyperprior on $\omega$ to formulate a Beta-Bernoulli prior, i.e. $\omega\sim\text{Beta}(a_{\omega},b_{\omega})$. The choice of $a_{\omega}$ and $b_{\omega}$ incorporates the prior belief that a certain percentage of taxa are discriminatory. 

We further specify a log-link function to integrate the covariates into the modeling construction for each \bch normalized \ech abundance:
\begin{equation}\label{cov}
	\begin{cases}\begin{array}{ll}
			\log\alpha_{ij0}\quad=\quad\mu_{0j}+\bm{x}_i\boldsymbol{\beta}_j^T & \text{if}~ \gamma_j =0\\ 
			\log\alpha_{ijk}\quad=\quad\mu_{0j}+\mu_{kj}+ \bm{x}_i\boldsymbol{\beta}_j^T & \text{if}~ \gamma_j  =1 \text{ and } z_i=k
		\end{array}
	\end{cases},
\end{equation}
where $\mu_{0j}$ is a feature-specific baseline parameter for taxon $j$. Note that $\exp(\mu_{0j})$'s can be considered as scaling factors adjusting for feature-specific levels across all samples. The group-specific parameter $\mu_{kj}$ captures the baseline shift between the $k$-th group and the reference group. We set $\mu_{kj}=0$ if the $k$-th group is the reference group to avoid identifiability problems arising from the sum of the components. $\bm{x}_i$, the $i$-th row of covariate matrix $\bm{X}$, contains all the covariate measurements for sample $i$. Here, $\boldsymbol{\beta}_j$ is a $1$-by-$R$ vector, with each element $\beta_{rj}$ modeling the global effect of the $r$-th covariate on the observed counts for the $j$-th taxon. In practice, not all of the covariates are related to the abundance of a taxon. Therefore, we allow different sets of covariates to affect different taxa by specifying a \textit{spike-and-slab} prior \citep{brown1998multivariate,ishwaran2005spike} as
$	\beta_{rj} \sim (1-\delta_{rj})I(\beta_{rj} = 0)+\delta_{rj} N(0,\sigma_{\beta j}^2)$,
where $\delta_{rj}=1$ indicates the $r$-th covariate is associated with the \bch normalized \ech abundance for the $j$-th feature, and $\delta_{rj}=0$ otherwise. This modeling approach allows us to identify significant covariate-taxa associations, via the selection of the nonzero $\beta_{rj}$ coefficients, for all discriminatory and non-discriminatory taxonomic features. We complete the model by setting $\mu_{0j} \sim  N(0,\sigma_{0 j}^2)$, $\mu_{kj} \sim  N(0,\sigma_{\mu  j}^2)$, and $\delta_{rj} \sim \text{Beta-Bernoulli}(a_p,b_p)$. Letting $\sigma_{0 j}^2 = 10^2$ for all $j$ yields a vague prior for the feature-specific baseline parameter. An inverse-gamma (IG) hyperprior $\text{IG}(a,b)$ is shared by $\sigma_{\mu  j}^2$ and $\sigma_{\beta j}^2$.
\vspace*{-1cm}

\subsection{Size factor estimation}\label{size factor}
The parameterization of the negative binomial mean, as shown in Equation (\ref{intergrate}), is a product of the size factor and the \bch normalized abundance\ech. It is typical to normalize the size factor first to ensure model identifiability. Hence, the plug-in estimator (equivalent to a point-mass prior) of $s_i$ is adopted to facilitate the inference based on the \bch normalized abundance \ech $\alpha_{ij}$ as shown in  Equation (\ref{cov}). The plug-in estimators can be calculated from the observed count matrix $\bm{Y}$. \bch There have been a number of proposals to estimate the size factors in the context of RNA-seq data analyses. Both \cite{witten} and \cite{li2017bayesian} conducted a comprehensive literature review. However, the assumptions of many existing methods for RNA-seq are likely not appropriate for highly diverse microbial environments \citep{Weiss2017}. \ech A so-called cumulative sum scaling (CSS) method has been developed by \cite{paulson2013differential} as $\hat{s}_i^\text{CSS} \propto  \sum_{j=1}^{p}y_{ij} I(y_{ij}\leq q_i^{l_{\text{CSS}}})$, where the default value of $l_{\text{CSS}}$ is $50$. CSS can be viewed as an adaptive extension of \cite{bullard2010evaluation}, and it is better suited for microbiome data. Moreover, a new normalization method named geometric mean of pairwise ratios (GMPR) has been proposed by \cite{chen2018gmpr}, aiming to handle the zero-inflated sequencing data. GMPR calculates the size factor $s_i$ based on the median count ratio of nonzero counts between the $i$-th sample and the remaining samples. It has been shown to be robust to differential and outlier OTUs. Combining this with some constraints such as $\sum_{i=1}^{n} \log\hat{s_i} = 0$ (i.e. $\prod_{i=1}^{n} \hat{s_i} = 1$), we are able to obtain a set of identifiable values. \bch In this paper, both CSS and GMPR are considered. \ech 
 
 
\section{Model Fitting and Posterior Inference}\label{fitting}
Our model space consists of $\left (\bm{R}, \boldsymbol{\phi}, \boldsymbol{\mu}_0, \bm{M}, \bm{B},  \boldsymbol{\gamma},  \boldsymbol{\Delta}, \omega, \pi \right )$ with the extra zero indicators $\bm{R} = (r_{ij},i=1,\ldots,n,j=1,\ldots,p)$, the dispersion parameters $ \boldsymbol{\phi} = (\phi_j, j=1,\ldots,p)$, the feature-specific baselines $ \boldsymbol{\mu}_0 = (\mu_{0j},j=1,\ldots,p)$, the group-specific baselines  $\bm{M} = (\mu_{kj},k=1,\ldots,K,j=1,\ldots,p)$, the covariate effects $\bm{B} = (\beta_{rj}, r = 1,\ldots,R,j=1,\ldots,p)$, the discriminatory taxa indicators $ \boldsymbol{\gamma} = (\gamma_j, j=1,\ldots,p)$, and the association indicators $ \boldsymbol{\Delta} = (\delta_{rj},r= 1,\ldots,R, j=1,\ldots,p)$. We explore the posterior distribution via a Markov chain Monte Carlo (MCMC) algorithm based on stochastic search variable selection with within-model updates \citep{savitsky2010spiked}. Full details can be found in the supplement. 

We are interested in distinguishing taxa that are differentially abundant among different groups, via $\boldsymbol{\gamma}$, as well as their associations with covariates, via $\boldsymbol{\Delta}$. One way to summarize the posterior distributions of these binary parameters is via the marginal posterior probability of inclusion (PPI). Suppose $t=1,\ldots,T$ index the MCMC iterations after burn-in. Then PPI of each $\gamma_j$ and $\delta_{rj}$ can be written as $\text{PPI}(\gamma_j) = \frac{1}{T}\sum_{t=1}^{T} \gamma_j^{(t)} \text{ and } \text{PPI}(\delta_{rj}) = \frac{1}{T}\sum_{t=1}^{T} \delta_{rj}^{(t)}$, respectively. Subsequently, important features and covariates can be selected based on a given PPI threshold. Following \cite{newton2004}, we choose a threshold that controls the Bayesian FDR. Specifically, we solve the following equations to determine the thresholds: $\text{FDR}_{\bm{\gamma}}(c_\gamma) = \frac{\sum_{j=1}^{p}(1-\text{PPI}({\gamma_j}))\text{I}(1-\text{PPI}(\gamma_j)<c_\gamma)}{\sum_{j=1}^{p}\text{I}(1-\text{PPI}(\gamma_j)<c_\gamma)},~
\text{FDR}_{\bm{\Delta}}(c_\delta) = \frac{\sum_{r = 1}^{R}\sum_{j=1}^{p}(1-\text{PPI}({\delta_{rj})})\text{I}(1-\text{PPI}(\delta_{rj}))<c_\delta)}{\sum_{r = 1}^{R}\sum_{j=1}^{p}\text{I}(1-\text{PPI}(\delta_{rj})<c_\delta)}$, where $\I(\cdot)$ is an indicator function. A well-accepted setting is to set both $ \text{FDR}_{\bm{\gamma}}$ and $\text{FDR}_{\bm{\Delta}}$ equal to $0.05$, which corresponds to an expected FDR of $5\%$.

\section{Simulation Study}\label{simulation}
In this section, we evaluated the proposed model using simulated data. In particular, we considered two methods (CSS and GMPR) introduced in Section \ref{size factor} for estimating the size factor $s_i$'s. We also compared our model with other existing methods described in the prior microbiome studies. In order to mimic metagenome sequencing data from real data applications (Section \ref{realdata}), we chose the parameters as follows: we set $n$ \ech samples for $K = 2$ groups with balanced group size $n_1 = n_2 = n/2$. We chose a large number of candidate features by setting the number of taxa $p = 300$, and randomly selected 20 true discriminant features to evaluate our model performance. Each row of $\bm{Y}$, denoted as $\bm{y}_i$, was generated from a Dirichlet-Multinomial distribution as described in \cite{wadsworth2017integrative}. For $i = 1, \ldots, n$, we let $\bm{y}_i \sim \text{Multinomial}(N_i, ~\boldsymbol{\pi}_i)$ with the row sum $N_i \sim \text{Discrete~Uniform}(2\times10^7,6 \times 10^7)$ and $\boldsymbol{\pi}_i = (\pi_{i1}, \ldots, \pi_{ip}) \sim \text{Dirichlet}(\bm{a}_i)$. We further incorporated the feature and covariate effects through $\bm{a}_i = (a_{i1}, \ldots, a_{ip})$ by setting $a_{ij} = \exp(a_{ij}^*)$ with $a_{ij}^* \sim  \text{Normal}(\mu_{0j} + \mu_{kj} + \bm{x}_i \boldsymbol{\beta}_j^T, ~\sigma_e^2)$. Here, a larger value of $\sigma_e^2$ corresponded to a higher noise level. Compared with Equation (\ref{intergrate}), this data generating process is different from the assumption of the proposed model. We set $\mu_{0j} \sim \text{Uniform}(8,10)$, $\mu_{1j} = 0$ for all $j$ and $\mu_{2j} = \pm  2$ for all selected discriminating features and 0 otherwise. Then for the covariate effects, we first obtained the covariate matrix $\bm{X}_{n \times R}$ by sampling each row $\bm{x}_i$ from the covariate matrix of the liver cirrhosis study in Section \ref{Liver cirrhosis} (with $n = 237$ and $R = 7$). In particular, we sampled $n/2$ covariate records from healthy and disease groups respectively. For each taxon $j$, we then randomly selected $m \in \{0,2,4,6\}$ out of $R$ covariates and let the corresponding $\beta_{rj} \sim \pm \text{Uniform}(0.5, 1)$ while setting the rest $\beta_{rj} = 0$.  Lastly, we randomly set $\pi_0np$ counts in $\bm{Y}$ to be zeros to mimic the zero-inflation in the real data. To summarize, we varied the following settings in order to comprehensively examine the model performance: 1) sample size per group $n / 2=10$ or $30$; 2) noise level $\sigma_e=0.5$, $1.0$, or $1.5$; 3) zero proportion $\pi_0=30\%$, $40\%$, or $70\%$. In the main text, we present the results obtained from the simulated datasets that $n/2 = 30, \sigma_e^2 = 1$, and $\pi_0 = 40\%$, and the remaining results can be found in Section S2 in the supplement. 

The hyperparameters were specified using the following default settings. For the binary variables with Beta-Bernoulli priors $\gamma_j \sim \text{Beta-Bernoulli}(a_{\omega},b_{\omega}),~ \delta_{rj}\sim \text{Beta-Bernoulli}(a_{p},b_{p})$ and $r_{ij}\sim \text{Beta-Bernoulli}(a_{\pi},b_{\pi})$, we set $a_{\omega} = 0.2, ~b_{\omega} = 1.8, ~a_p = 0.4,~\text{and}~b_p = 0.6$, which means that $10\%$ of the taxa are expected to be discriminant features, and $20\%$ of the covariate coefficients to be nonzero. We chose $a_{\pi} = b_{\pi} = 1$ assuming that about half of the zeros are truly missing. For the dispersion parameter with Ga($a_{\phi},b_{\phi})$ prior, we set $a_{\phi} =  1,~b_{\phi} = 0.01$ to obtain a vague gamma prior with mean of 100 and variance of 10,000. Next, we specified a flat prior $\text{IG}(a = 2, ~b = 10)$ for the variance term $\sigma_{\mu j}^2$ and $\sigma_{\beta j}^2$. The sensitivity analysis reported in the Section S3.2 in the supplement contains more details on the choice of $a$ and $b$. When implementing our model on a dataset, we ran four independent chains with different starting points where each feature or covariate was randomly initialized to have $\gamma_j = 1~\text{or}~0, \delta_{rj} = 1~\text{or}~0$. We set $20,000$ iterations as the default and discarded the first half as burn-in. To assess the concordance between four chains, we looked at all the  pairwise correlation coefficients between the marginal
PPI of $\boldsymbol{\gamma}$ and $\boldsymbol{\Delta}$. As mentioned by \cite{stingo2013integrative}, high values of correlation suggest that MCMC chains are run for a satisfactory number of iterations. After ensuring convergence, we assessed our model performance based on the averaged result over four chains.

Our goal was to identify the discriminating features (e.g. taxa) and the significant feature-covariate associations (i.e. all nonzero $\gamma_j$ and $\delta_{rj}$ in our model). We thus obtained the PPI for all $\gamma_j$ and $\delta_{rj}$, and visualized the accuracy in feature selection using the receiving operating characteristic (ROC) curve. We also computed the false positive rate when all feature-covariate associations were zero. \bch We further considered two types of competitors for model comparison. The first type, similar to the proposed model, can simultaneously identify discriminating features and detect the feature-covariate associations. Here, we compared with the multivariate zero-inflated logistic-normal (MZILN) regression model proposed by \cite{li2018conditional}. The MZILN model treats the sample allocation vector as an observed covariate for each sample. Therefore we combined the group label with other observed covariates to create a new covariate matrix, and the MZILN model gave a regularized estimation of the regression coefficient between each feature and covariate. The selected discriminating features and feature-covariate associations corresponded to the nonzero coefficient estimations. The second type of method achieves the same goal in two separate stages. The first stage consists of four methods to select discriminating features based on $p$-values, including the Wilcoxon rank-sum test (Wilcoxon test) and three differential expression analysis methods implemented by the \texttt{R} packages \texttt{metagenomeSeq} \citep{paulson2013differential}, \texttt{edgeR} \citep{robinson2010edger} and \texttt{limma} \citep{ritchie2015limma}. Specifically, \texttt{metagenomeSeq} assumes a zero-inflated Gaussian model, \ech \verb|edgeR| models count data using a negative binomial distribution, and \verb|limma| adopts a linear model for the log-transformed count data. Then, the discriminating features were selected to be those with BH \citep{benjamini1995controlling} adjusted $p$-values smaller than $0.05$. To make a head to head comparison in the first stage, we also included a simplified version of the ZINB model by excluding the covariate term $\bm{x}_i\boldsymbol{\beta}_j^T$ in Equation (\ref{cov}). In the second stage, we considered the following feature selection strategies for each $p$-value based method. They are: 1) correlation test, 2) lasso regression, 3) random forest, and 4) multivariate linear regression. We centered the selected discriminating features by group, and the rest across all samples. For the correlation test, the Pearson correlation coefficients were calculated between the log scaled compositional data and the covariate measurements for each outcome group. Next, a Fisher z-transformation \citep{fisher1915frequency} was applied to obtain the $p$-values for testing the significance of correlation. For lasso regression, we calculated the true positive rates and the false positive rates with respect to a range of lasso penalty. For the last two, we fitted a random forest model or a multivariate linear regression model between each feature and the covariate matrix $X$, which yielded variable importance measures or $p$-values. In all, we have four choices in the first stage \{Wilcoxon test, \texttt{metagenomeSeq}, \texttt{edgeR}, \texttt{limma}\} and four choices in the second stage \{correlation test, lasso regression, random forest, multivariate linear regression\}, with $4 \times 4  = 16$ choices in total. \bch For clear visualization of the result, we excluded \texttt{limma} in the second stage due to its relatively inferior performance in the first stage. We also dropped random forest and linear regression since they showed similar performance as the lasso regression. \ech Besides, all the $p$-values generated using different methods were adjusted using the BH method to control the FDR.

\begin{figure}
	\centering
	\begin{tabular}{@{}c@{}}
		\subfloat[][ROC curves for $\bm{\gamma}$ (left) and false positive rates for $\bm{\Delta}$ (right) with all covariate coefficients $\beta_{rj}$ as $0$ ]{\includegraphics[width=0.7\textwidth]{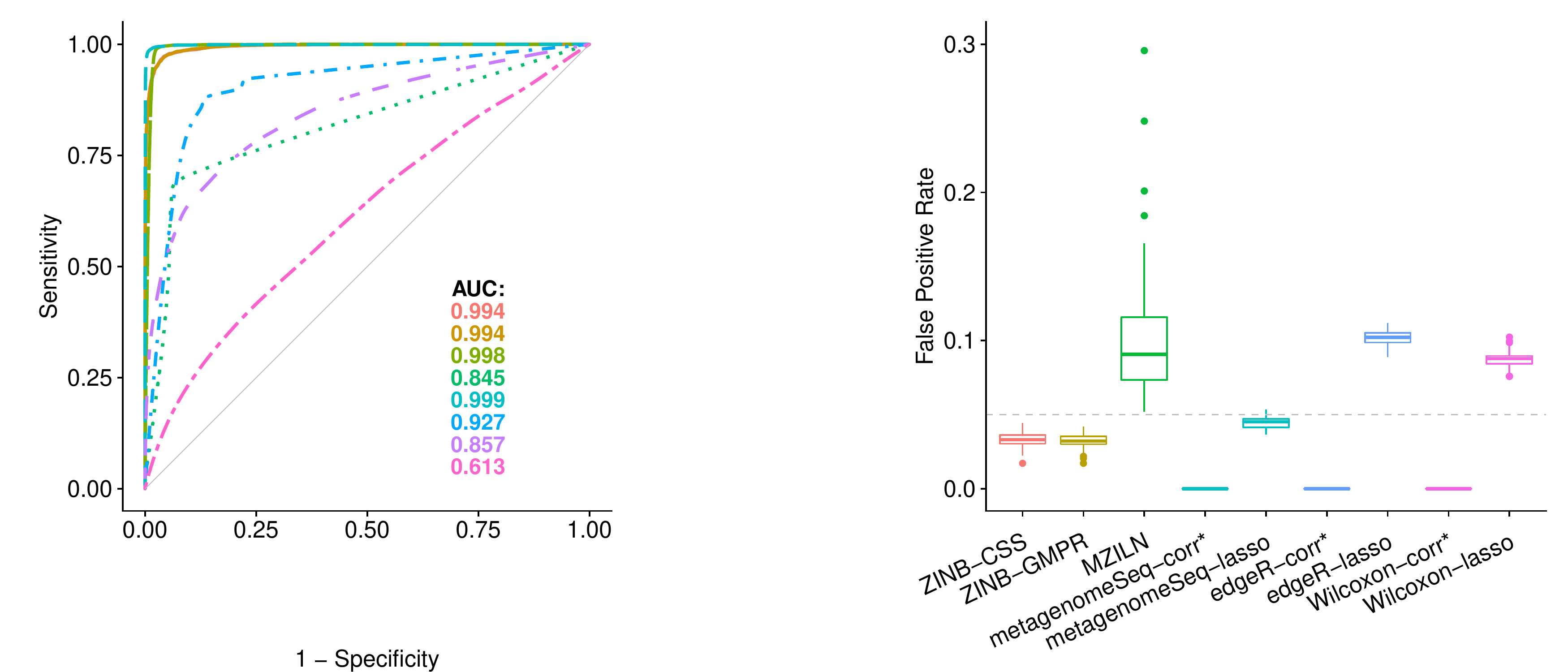}\label{simu0}} \\
		\subfloat[][ROC curves for $\bm{\gamma}$ (left) and $\bm{\Delta}$ (right) with a number of nonzero covariate coefficients $\beta_{rj}$ of 2]{\includegraphics[width=0.6\textwidth]{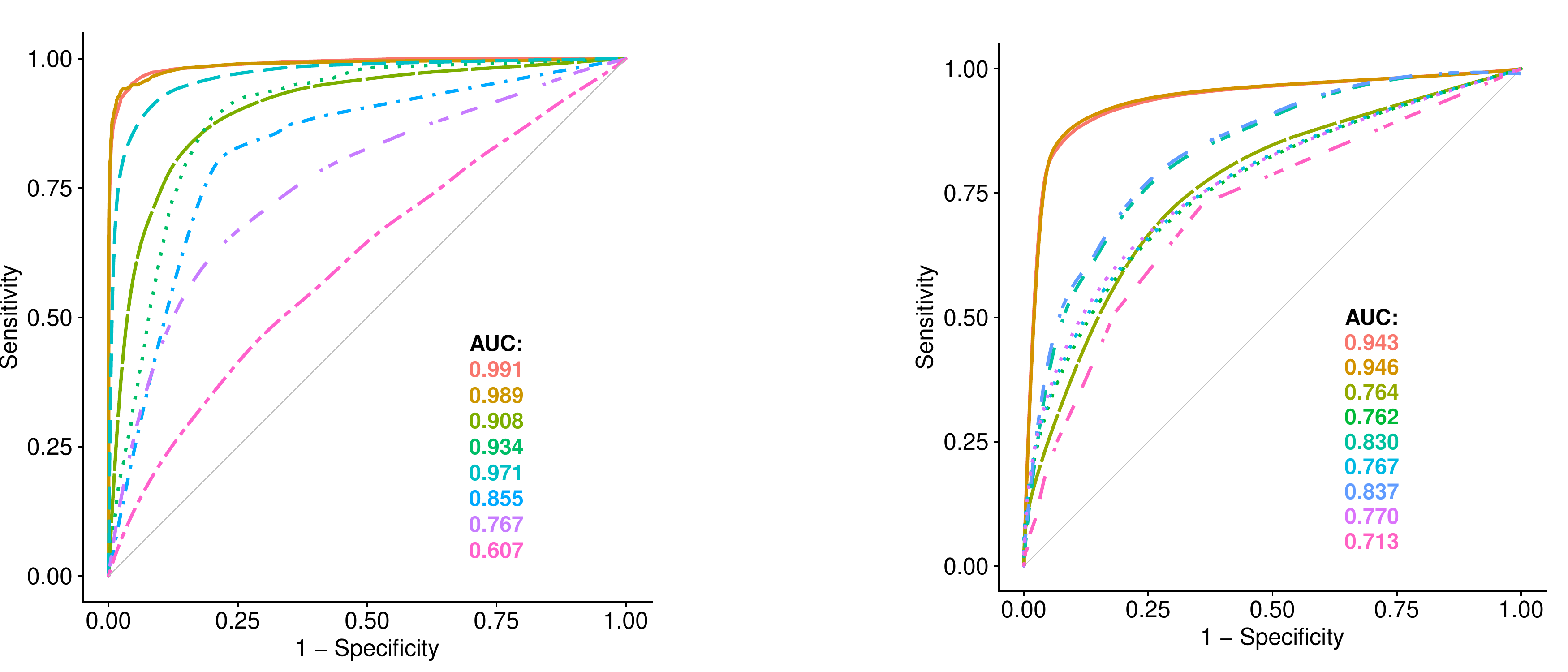}\label{simu2}} \\
		\subfloat[][ROC curves for $\bm{\gamma}$ (left) and $\bm{\Delta}$ (right) with a number of nonzero covariate coefficients $\beta_{rj}$ of 4]{\includegraphics[width=0.6\textwidth]{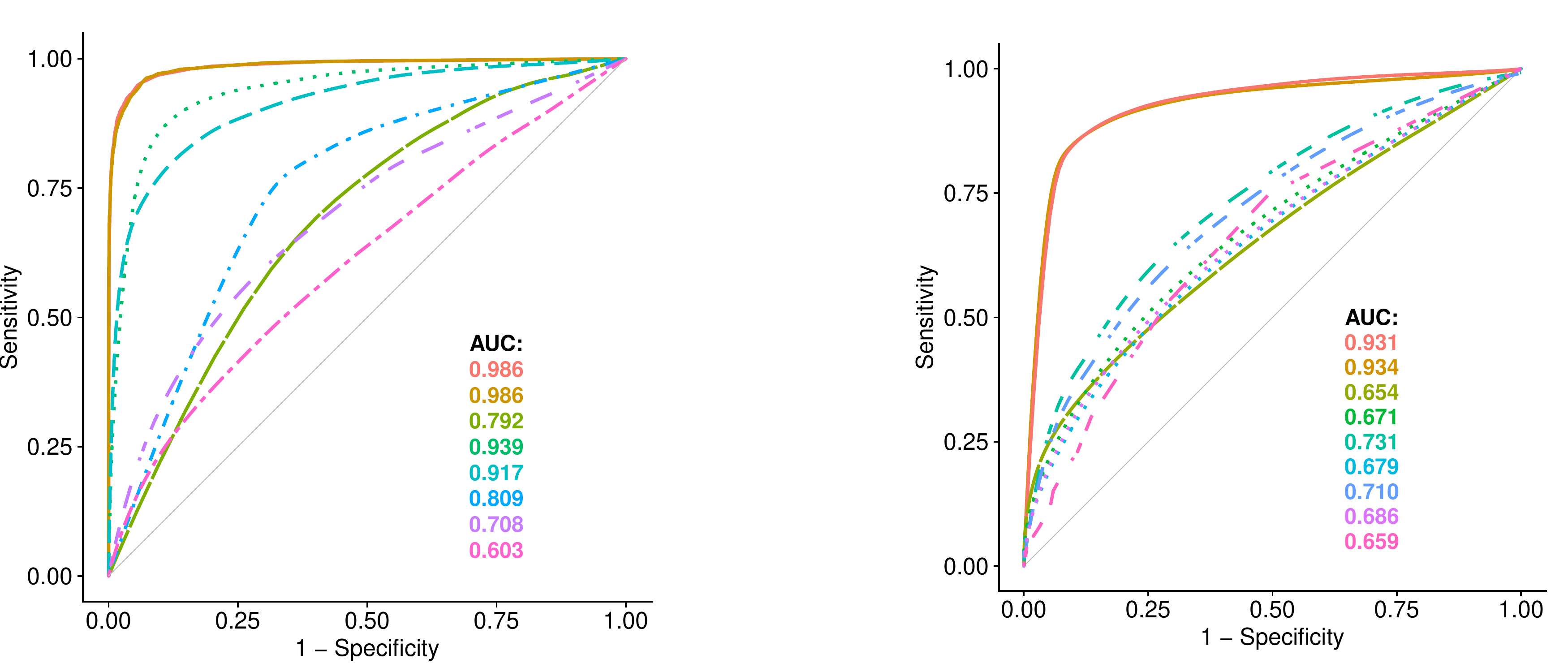}\label{simu4}} \\
		\subfloat[][ROC curves for $\bm{\gamma}$ (left) and $\bm{\Delta}$ (right) with a number of nonzero covariate coefficients $\beta_{rj}$ of 6]{\includegraphics[width=0.6\textwidth]{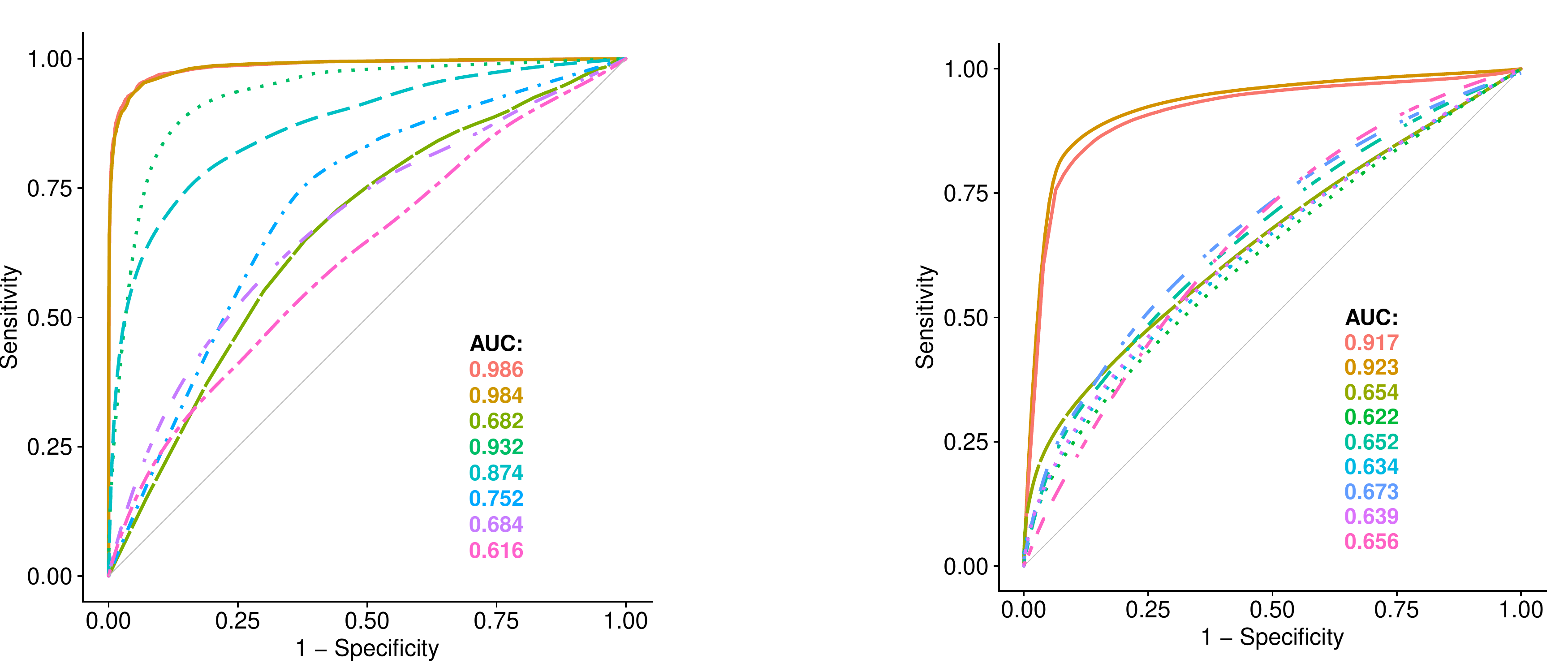}\label{simu6}} 
	\end{tabular}
	\begin{tabular}{@{}cc@{}}
		\centering
		\subfloat{\includegraphics[width=0.5\textwidth]{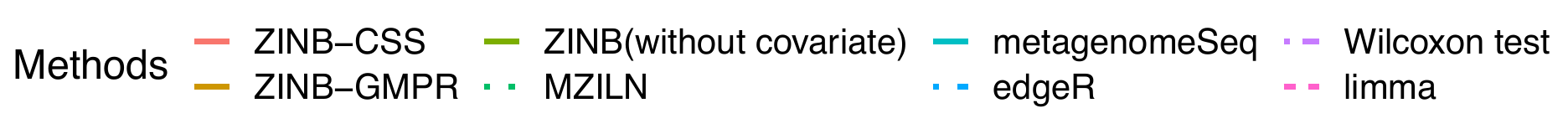}} & 
		\subfloat{\includegraphics[width=0.5\textwidth]{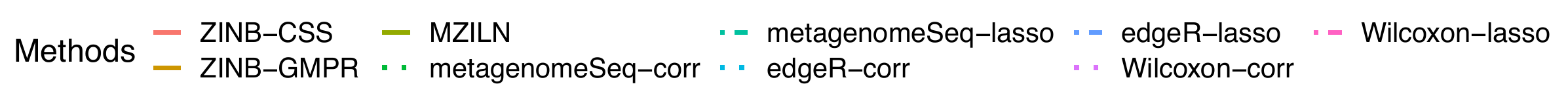}}
	\end{tabular}
	\caption[]{\bch Averaged ROC curves for the discriminating feature indicator $\bm{\gamma}$ (left) and the feature-covariate association indicator $\bm{\Delta}$ (right) with respect to different numbers of nonzero covariate coefficients, i.e. (a) $0$, (b) $2$, (c) $4$, and (d) $6$ out of $7$, over $100$ replicates in each scenario.\\
	$^*$The correlation-based methods showed low false positive rates in the case where there is no truly nonzero covariate coefficients. \ech}
	\label{simu_res}
\end{figure}

For each of the four scenarios, Figure \ref{simu_res} compares the model performance through the averaged ROC curve over 100 simulated datasets. We also include the area under curve (AUC) for each approach. For detecting discriminating features, the proposed method  consistently shows high AUC ($ > 0.98$) across all scenarios, and similar results for capturing the feature-covariate associations (AUC $ > 0.90$). Moreover, the proposed method maintains a low FDR even when all $\beta_{rj}$ are 0. The correlation-based method shows low false positive rates in the case where the true number of contributing covariate is 0, but has low power when $\{2, 4, 6\}$ out of $7$ covariates have nonzero contribution. \bch In addition, the proposed model achieves the highest true positive fraction under a fixed small value of FDR in all scenarios, with the MZILN model and \texttt{metagenomeSeq} performing the second and third best in estimating the discriminating feature indicator $\boldsymbol{\gamma}$ (shown in the left column of Figure \ref{simu_res}, Figures S3-S5). We also noticed that the MZILN model could not outperform the two-stage methods in estimating the feature-covariate association indicator $\boldsymbol{\Delta}$. The above conclusions hold for using either CSS or GMPR to estimate the plug-in size factors. To test if our model is robust to the choice of size factor estimation methods, we further conducted a sensitivity analysis in Section S3.1 in the supplement. The result, as shown in Figure S6, suggests that our model is considerably robust to the choice of different normalization methods, while CSS and GMPR have a marginal performance improvement.  \ech Furthermore, we reach the same conclusion with varying group sizes, log-scale noise levels, and zero proportions. In particular, the proposed ZINB model is robust to a larger amount of extra zeros. Either decreasing the group size or increasing the noise level hampers the performance of all the methods. Nevertheless, the ZINB model still consistently outperforms the alternative approaches in estimating $\bm{\gamma}$ and $\bm{\Delta}$. Results are summarized in Figures S3-S5 in the supplement.

\section{Feature-Covariate Association Analysis}\label{association-analysis}
In this section, we demonstrate that our model can estimate the association between a taxonomic feature and a covariate by adjusting for the remaining confounders. As a comparison, current approaches rely on correlation analysis between the pairwise microbiome and covariates. Specifically, those analyses converted each observed taxonomic count to a fraction (or termed percentages, intensities) by sample. Next the Pearson correlation coefficients were calculated between the log scaled fractions and the covariate measurements for each outcome group. Lastly, a Fisher z-transformation \citep{fisher1915frequency} was applied to obtain the $p$-values for testing the significance of correlation.

Our model constructs a regression framework to quantify the relationship between the \bch normalized \ech abundance $\alpha_{ijk}$ and covariates through the Equation (\ref{cov}).
Based on Equation (\ref{cov}), given a feature $j$ and a covariate $r$ of interest, we first normalized the observed abundance using CSS and performed logarithmic transformation. Next, to calculate $x_{ir}\hat{\beta}_{rj}$ and the group shift $\hat{\mu}_{kj}$, we subtracted the estimated feature-specific influence $\hat{\mu}_{0j}$ and other covariates' impact $\sum_{r'\neq r}x_{ir'}\hat{\beta}_{r'j}$ from the transformed abundance. Lastly, we could evaluate whether our model provided a reasonable estimation ($\hat{\beta}_{rj}$) of the feature-covariate association between covariate $r$ and the normalized and adjusted observations of feature $j$.

\begin{figure}
	\centering
	\begin{tabular}{@{}cc@{}}
		\subfloat[][Estimation of $\hat{\beta}_{1, 68}$ from the ZINB model ]{\includegraphics[width=.5\textwidth]{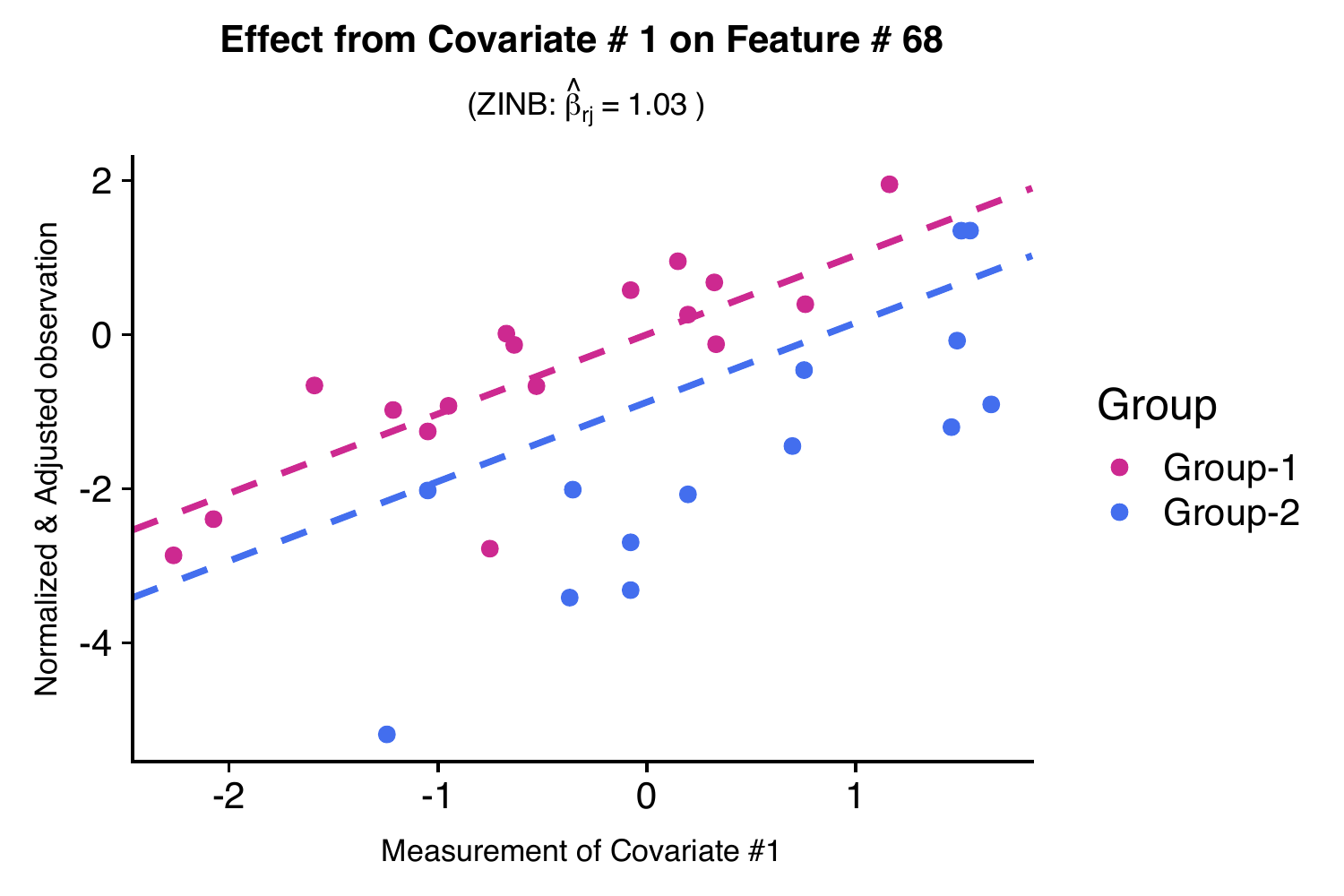}\label{feature-cov a}} & 
		\subfloat[][Estimation of $\hat{\beta}_{1, 127}$ from the ZINB model]{\includegraphics[width=.5\textwidth]{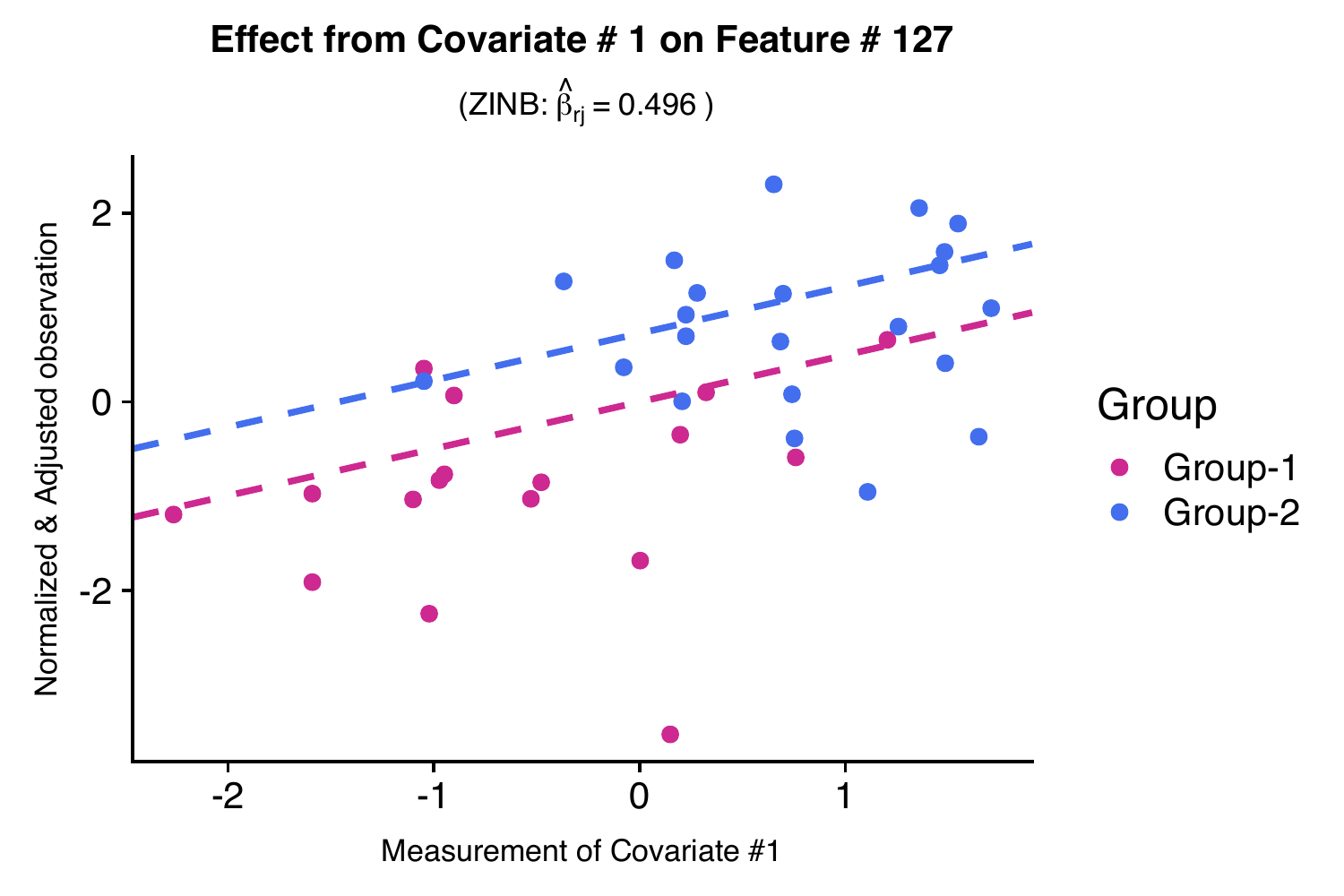}\label{feature-cov b}}\\
		\subfloat[][Pearson correlation coefficient between feature 68 and covariate 1(with corresponding $p$-value)]{\includegraphics[width=.5\textwidth]{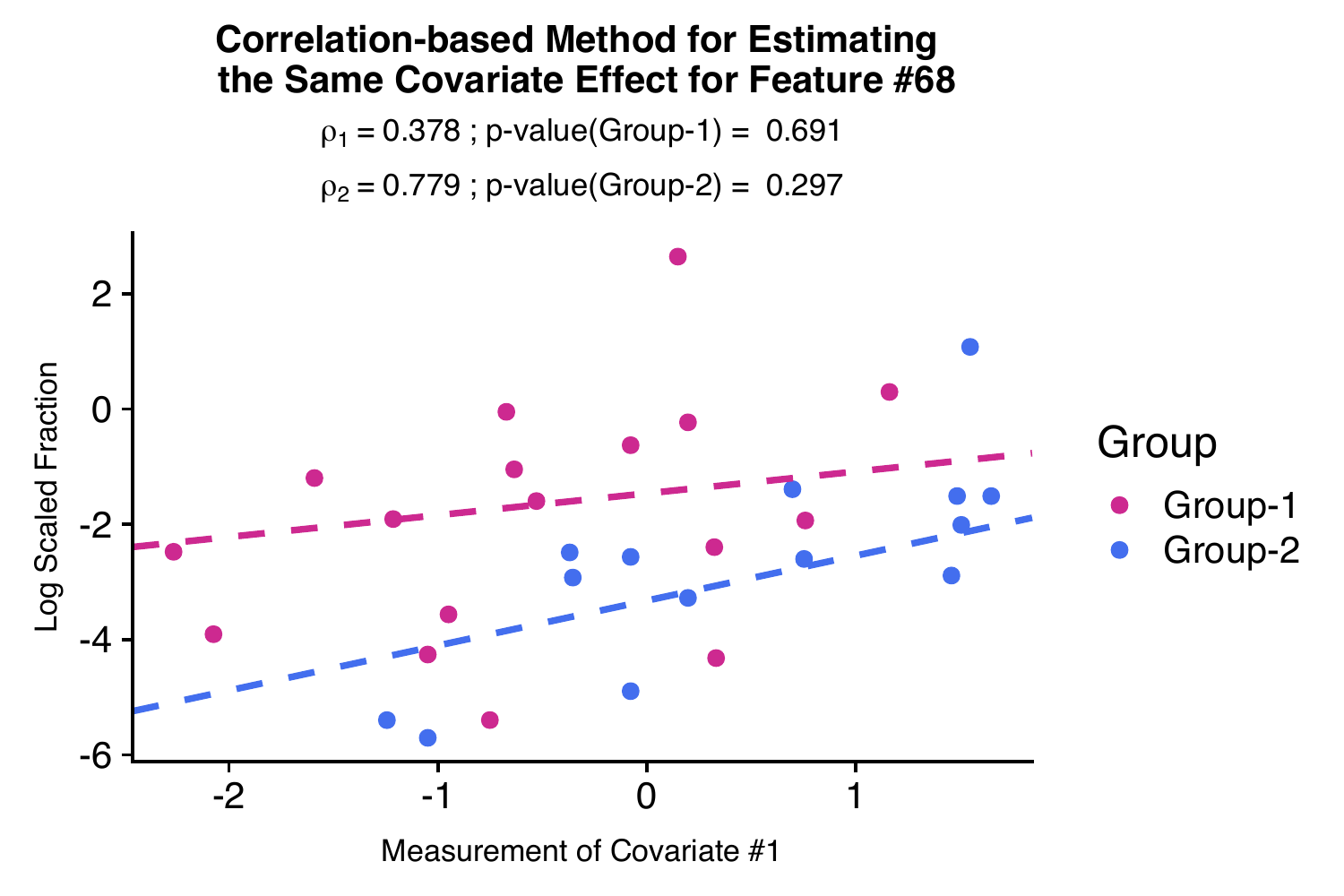}\label{feature-cov c}} &
		\subfloat[][Pearson correlation coefficient between feature 127 and covariate 1(with corresponding $p$-value)]{\includegraphics[width=.5\textwidth]{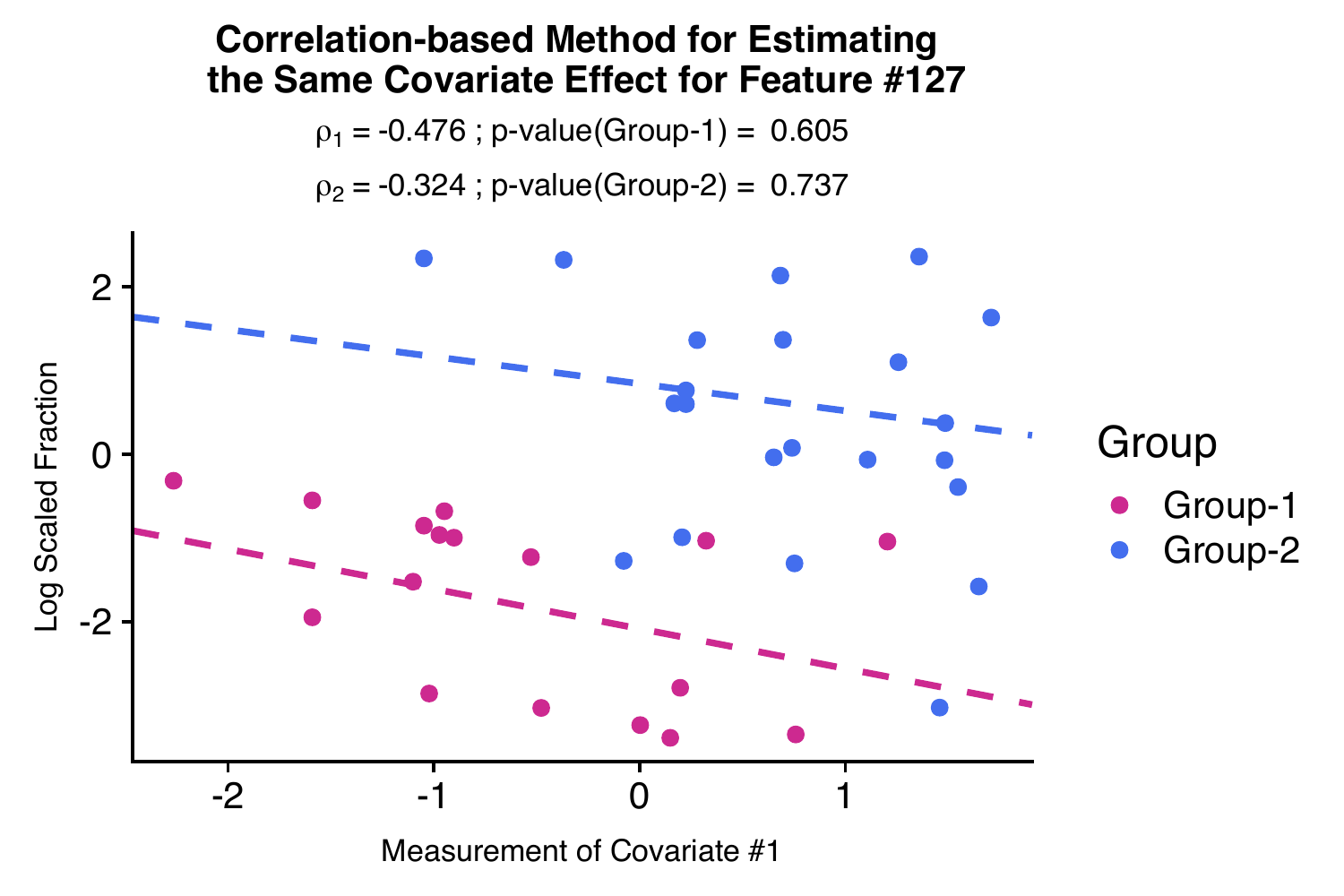}\label{feature-cov d}}
	\end{tabular}
	\caption[]{Feature-Covariate Association Analysis: comparison of the results given by the proposed method ((a) and (b)) and correlation-based method((c) and (d)) from the simulated dataset, where the two features shown (randomly selected for illustration) were truly discriminating with the covariate effect $\beta_{1, 68} > 0$ and $\beta_{1, 127} < 0$ by simulation. The proposed method provided a reasonable estimation ($\hat{\beta}_{rj}$) of the feature-covariate association.}
	\label{feature-cov}
\end{figure}

Here, we demonstrated the advantages of the proposed model in estimating the feature-covariate association over the correlation-based method through simulation. For each feature, we randomly selected four out of seven covariates to have nonzero linear effects on the latent abundances, and generated a simulated dataset following the description in Section \ref{simulation}. We kept the same prior and algorithm settings to obtain the estimations for all parameters of interest. We chose a $5\%$ Bayesian FDR for estimating $\boldsymbol{\Delta}$. Among all feature-covariate combinations, the proposed model achieved sensitivity and specificity rates of $82.9\%$ and $86.7\%$ respectively. We randomly chose several pairs of feature and covariate and compared the proposed method and the correlation-based method. Figure \ref{feature-cov} displays the results of 2 examples, where the true values of $\delta_{rj}$ were 1. The two dashed lines in Figure \ref{feature-cov}a or \ref{feature-cov}b have the same slope of $\hat{\beta}_{rj}$ as our estimated covariate effect. Both plots suggest that the proposed model is able to capture the feature-covariate relationship. Notice that we did not adjust for the group-specific effect. Hence the differences between two dashed lines represents the group-specific parameter $\hat{\mu}_{kj}$. These results illustrated the advantages in simultaneously detecting the discriminating features and quantifying the feature-covariate associations. Furthermore, we also validated that the proposed model had correctly captured the direction of covariate effects in both cases. Figure \ref{feature-cov}b and \ref{feature-cov}d show the results from the correlation-based model. The slope of each dashed line represents the Pearson correlation coefficient. However, there was no significant result as all $p$-values were greater than $0.05$, suggesting that the correlation test can be underpowered. The correlation-based method failed to isolate the covariate of interest from the confounders, and it might suggest a wrong direction of covariate effect, as shown in Figure \ref{feature-cov}d.

\section{Real data analysis}\label{realdata}
We applied the proposed model on two real data sets: one with hundreds of samples and the other with only 24 samples. Compared with the analysis methods used in the original publications, our model demonstrates better performance in detecting differentially abundant bacteria. In addition, our model supports adjusting for biologically meaningful covariates. When adjusting for the metabolic pathway quantities (or metabolites through metabolomics technology) as covariates, our model estimates the association between taxa and metabolism-related functions (or metabolites). 

\subsection{Liver cirrhosis dataset}\label{Liver cirrhosis}
Cirrhosis is a late-stage condition of scarring or fibrosis of the liver caused by liver disease such as hepatitis B, hepatitis C, and non-alcoholic fatty liver disease \citep{abubakar2015global}. The liver is connected to the gastrointestinal tract via the hepatic portal and bile secretion systems. Interestingly, distinct gut microbiota signatures have been associated with both early-stage liver diseases and end-stage liver cirrhosis \citep{garcia2004gut, yan2011enteric, benten2012gut}. We applied our model on a gut microbiome dataset from a liver cirrhosis study carried out by \cite{qin2014}. All metagenome sequenced samples were available from the NCBI Short Read Archive and the curated microbial abundance matrix was accessible from ExperimentHub \citep{pasolli2017accessible}. 

The full dataset includes 237 samples with their observed microbial abundance matrix $\bm{Y}$ profiled from the gut microbiome at the species taxonomic level. The study has two patient groups, including 114 healthy controls and 123 liver cirrhosis patients. We filtered out the taxa with extremely low abundance before the analysis as suggested in \cite{wadsworth2017integrative}. We obtained 528 taxa that had at least 2 observed counts in both groups for further analysis. As for the covariate information, we used MetaCyc, a collection of microbial pathways and enzymes involved in metabolism for an extensive amount of organisms \citep{caspi2007metacyc}. We incorporated the 529 MetaCyc pathway measurements for 237 individuals in the study, and reduced the high correlation among the pathways by average linkage clustering on their correlation matrix \citep{wadsworth2017integrative}. Specifically, we kept the pathway with the largest fold-change between two groups in each cluster, and decided the number of clusters such that the correlations between the resulting pathways were less than 0.5. Logarithmic transformation and normalization (zero mean and unit variance) were applied to the selected covariates to ensure the zero mean and unit standard deviation. After the pre-processing step, we had seven covariates representing metabolic functions.

In \cite{qin2014}, differential analyses were based on the Wilcoxon test and the $p$-values were corrected by the BH method. Although a stringent threshold of significance level (0.0001) was used, the authors discovered 79 differentially abundant species and had to restrictively report the 30 top candidates in each group. Figure \ref{cladogram1} is the cladogram of the discriminating taxa selected by different methods, with blue dots representing the results by \cite{qin2014} and red dots reported by the ZINB model. As suggested in our simulation study, these results may reflect a high FDR as covariate effects were not factored in the analysis. In addition, the Wilcoxon test cannot account for the pathway effects and thus the associations between bacteria and metabolic pathways were not identified.

We applied the proposed Bayesian ZINB model to \bch simultaneously \ech analyze the microbial abundance matrix of bacteria and their metabolic pathway abundance. We set a similar hyperparameter setting as discussed in Section \ref{simulation} by first specifying $ a_{\mu} = a_{\beta} = 2$ and $b_{\mu} = b_{\beta} = 10$. Next, we set $a_{\omega}, ~b_{\omega}, ~a_p, ~b_p,~a_{\phi}, ~b_{\phi}$ to be the same as their default values discussed in Section \ref{simulation}. We ran four independent Markov chains with different starting points. Each chain had 40,000 iterations with the first half discarded as burn-in. We checked the convergence visually and calculated the pairwise Pearson correlation for PPIs, which ranged from 0.988 to 0.994 for $\gamma$'s and from 0.982 to 0.989 for $\delta$'s. These concluded highly consistent results. Figure \ref{realPPI1} shows the PPIs for all 528 taxa, where the dashed line represents the threshold corresponding to an expected FDR of  0.05. We identified 19 differentially expressed taxa, the majority of which are more abundant in the liver cirrhosis group. 

\begin{figure}
	\centering
	\begin{tabular}{@{}cc@{}}
		\subfloat[][Liver cirrhosis study: heatmap of covariate effect for all taxa]{\includegraphics[width=.5\textwidth]{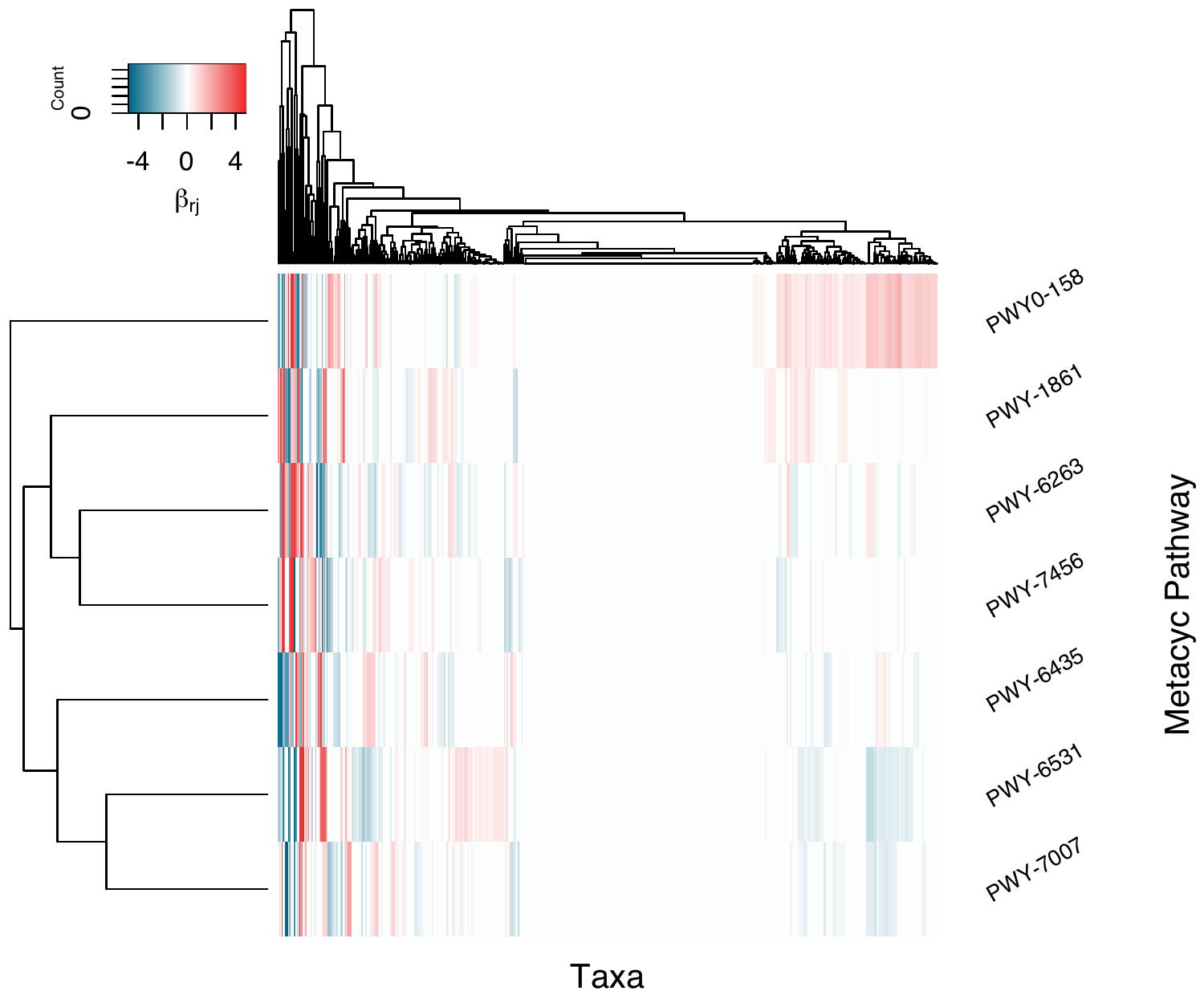}\label{heatmap1}}& 
		\subfloat[][Liver cirrhosis study: heatmap of covariate effect for selected discriminating taxa]{\includegraphics[width=.5\textwidth]{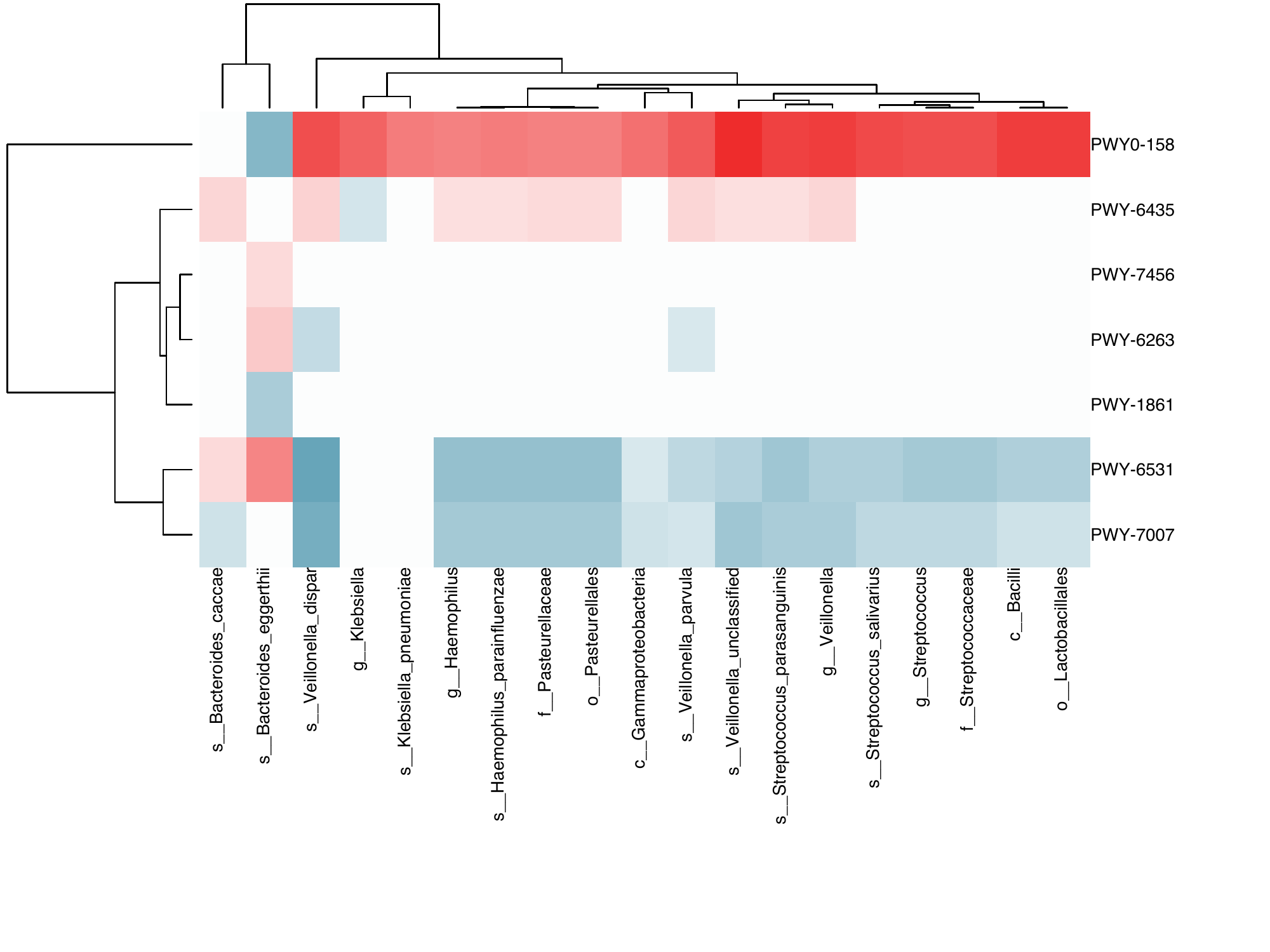}\label{heatmap12}}\\
			\subfloat[][Metastatic melanoma study: heatmap of covariate effect for all taxa]{\includegraphics[width=.5\textwidth]{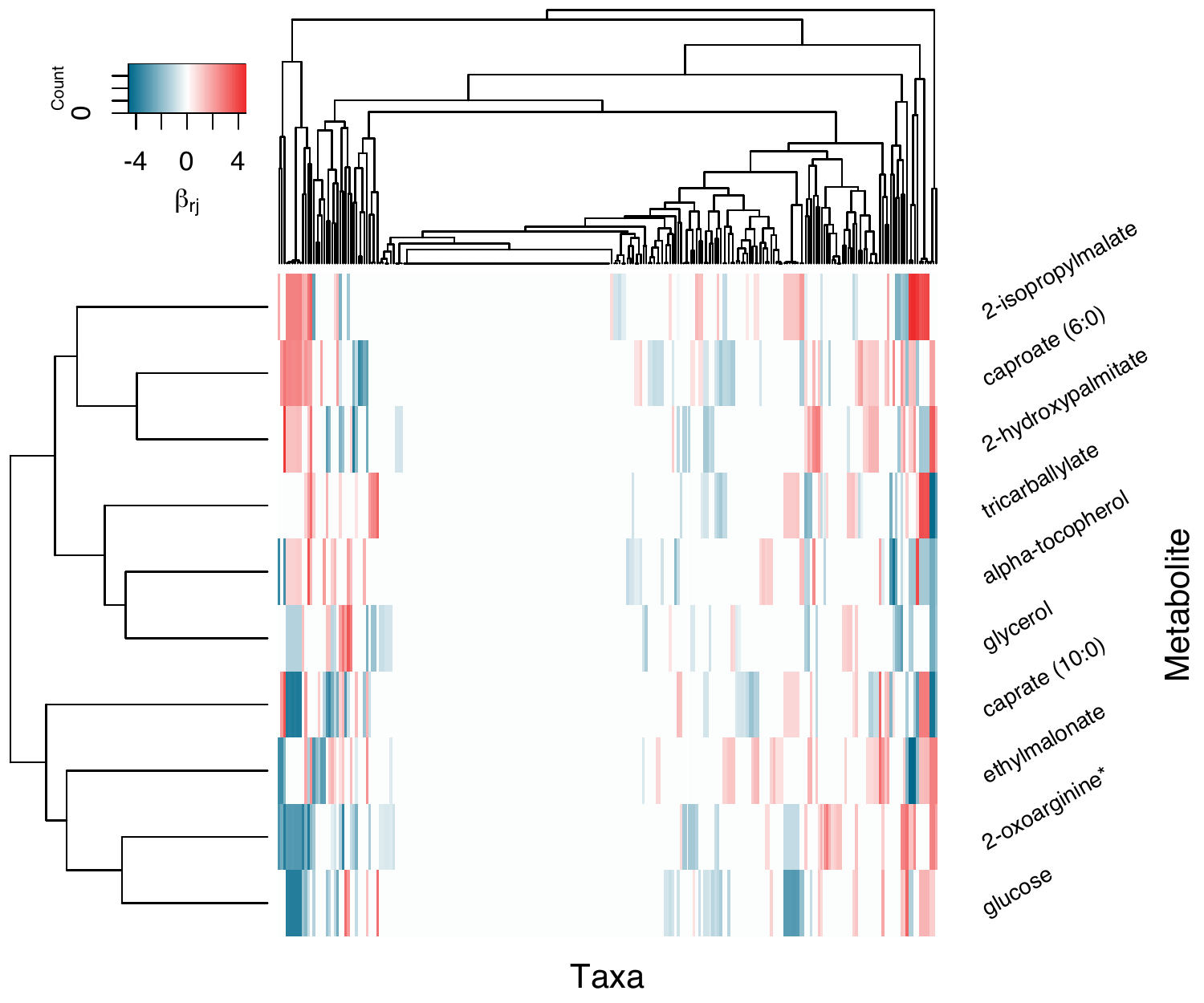}\label{heatmap2}} &
		\subfloat[][Metastatic melanoma study: heatmap of covariate effect for selected discriminating taxa]{\includegraphics[width=.5\textwidth]{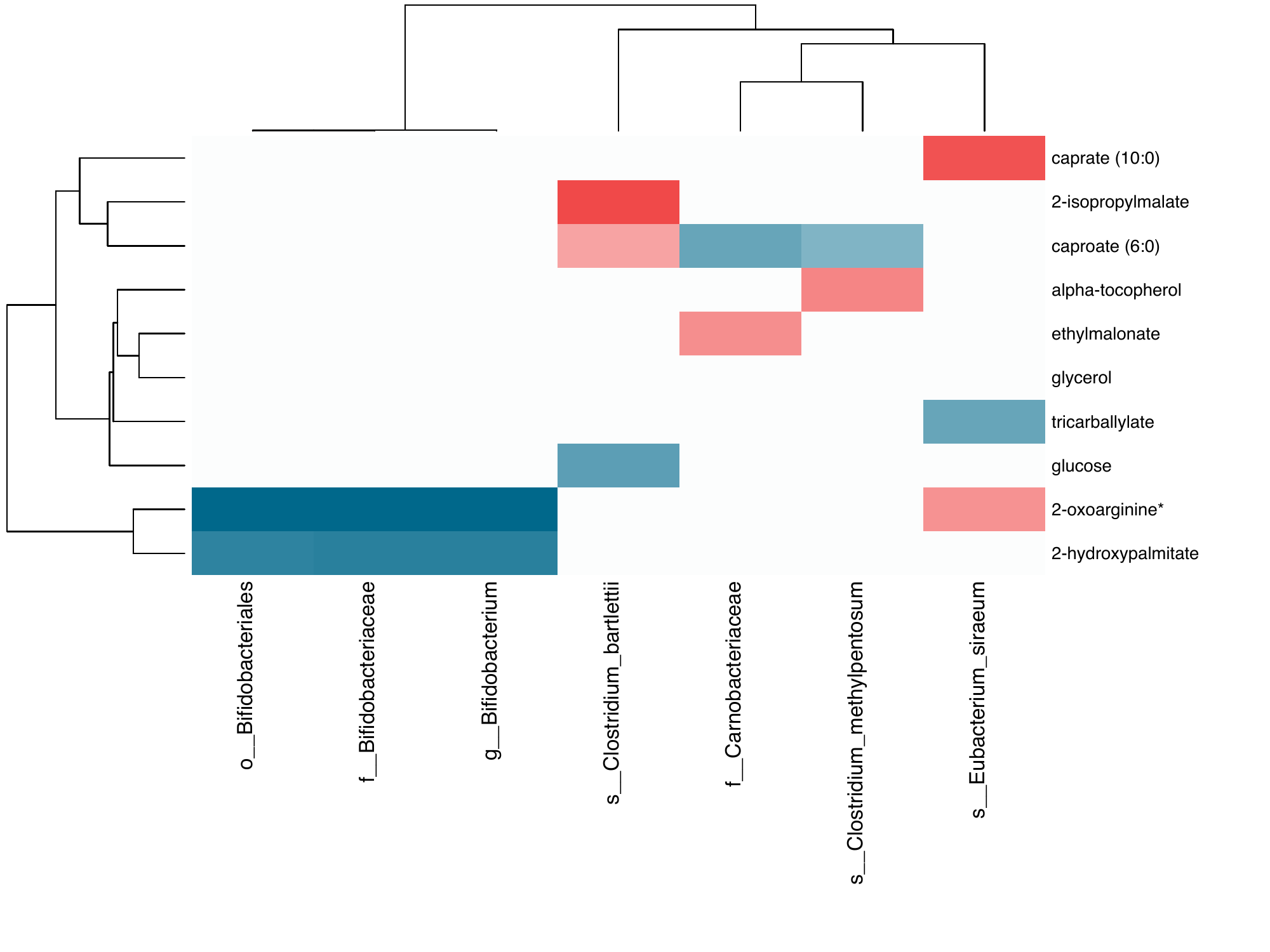}\label{heatmap22}}
	\end{tabular}
	\caption{Real Data Analysis: Heatmap showing the effect from covariates, the MetaCyc pathway abundances, in two studies. (a)(b), we use a liver cirrhosis dataset and show the effect between covariate effects and all microbiome, or differential abundant microbiome, respectively;  (c)(d), we use metastatic melanoma dataset and show the effect between covariate effects and all microbiome, or differential abundant microbiome, respectively;) }
	\label{heatmap}
\end{figure}

Figure \ref{CI1} shows the posterior mean of $\mu_{2j}$ for all identified discriminating taxa, and Table S1 in the supplement contains all the detailed parameter estimations for those taxa. Interestingly, two clear taxonomic branches are distinguished by our model (as indicated by red dots, Figure \ref{cladogram1}): the genera \textit{Veillonella} and \textit{Streptococcus}, both of which can originate from the oral cavity. Of note, oral commensal bacteria are able to colonize the distal intestinal tract in liver cirrhosis patients \citep{qin2014}, probably due to bile acid changes. \textit{Veillonella} spp. and \textit{Streptococcus} spp. have been identified as more abundant in patients with primary biliary cholangitis \citep{tang2017gut}, another hepatic disorder which shares pathophysiologic features with liver cirrhosis \citep{ridlon2013cirrhosis}. Figure \ref{heatmap1} and \ref{heatmap12} show the identified associations between microbiota and metabolic pathways. For example, L-alanine biosynthesis (PWY0-1061) is positively correlated with \textit{Veillonella}. Alanine is a gluconeogenesis precursors in liver metabolism, and increased alanine is thought to induce pyruvate kinase in \textit{Veillonella}. Thus, this connection between alanine synthesis and \textit{Veillonella} is intriguing and potentially novel, and biologic validation experiments might offer further clarification.

\subsection{Metastatic melanoma dataset}\label{Metastatic melanoma}
The proposed Bayesian ZINB model can perform integrative analysis of microbiome taxonomic data and other omics datasets. In this section, we applied this model to \bch simultaneously \ech analyze microbiome and metabolomics data from a study of advanced stage melanoma patients receiving immune checkpoint inhibitor therapy (ICT) \citep{frankel2017}. The data were collected using MSS and unbiased shotgun metabolomics. Here, we aim at identifying unique microbiome taxonomic and metabolomic signatures in those patients who responded favorably to ICT. 

A subset of patients in this study ($n = 24$) were treated with ipilimumab and nivolumab(IN), a combination therapy that has been shown to be more efficacious than therapy with anti-PD1 or anti-CTLA4 therapy alone. 16 patients responded to treatment and 8 patients had progression. We performed quality control steps on MSS reads and profiled them using MetaPhlAn \cite{segata2012metagenomic} as described in \citep{frankel2017}. We filtered out taxa with at most one observation in either patient group, which left $p = 248$ taxa from species to kingdom level. For the same fecal samples, we performed metabolomics profiling and quantified 1,901 patients' metabolite compounds as the covariate matrix $\bm{X}$. We are interested in statistically assessing how the biochemical volumes between patient groups are associated with bacteria burden or quantities. We adopted the same strategy mentioned in section \ref{Liver cirrhosis} to reduce the correlation between covariates, which resulted in a $24 \times 9$ matrix as the covariate matrix of $\bm{X}$.

\begin{figure}
	\centering
	\begin{tabular}{@{}cc@{}}
		\subfloat[][Liver cirrhosis study: plot of $\gamma$ PPI]{\includegraphics[width=.5\textwidth]{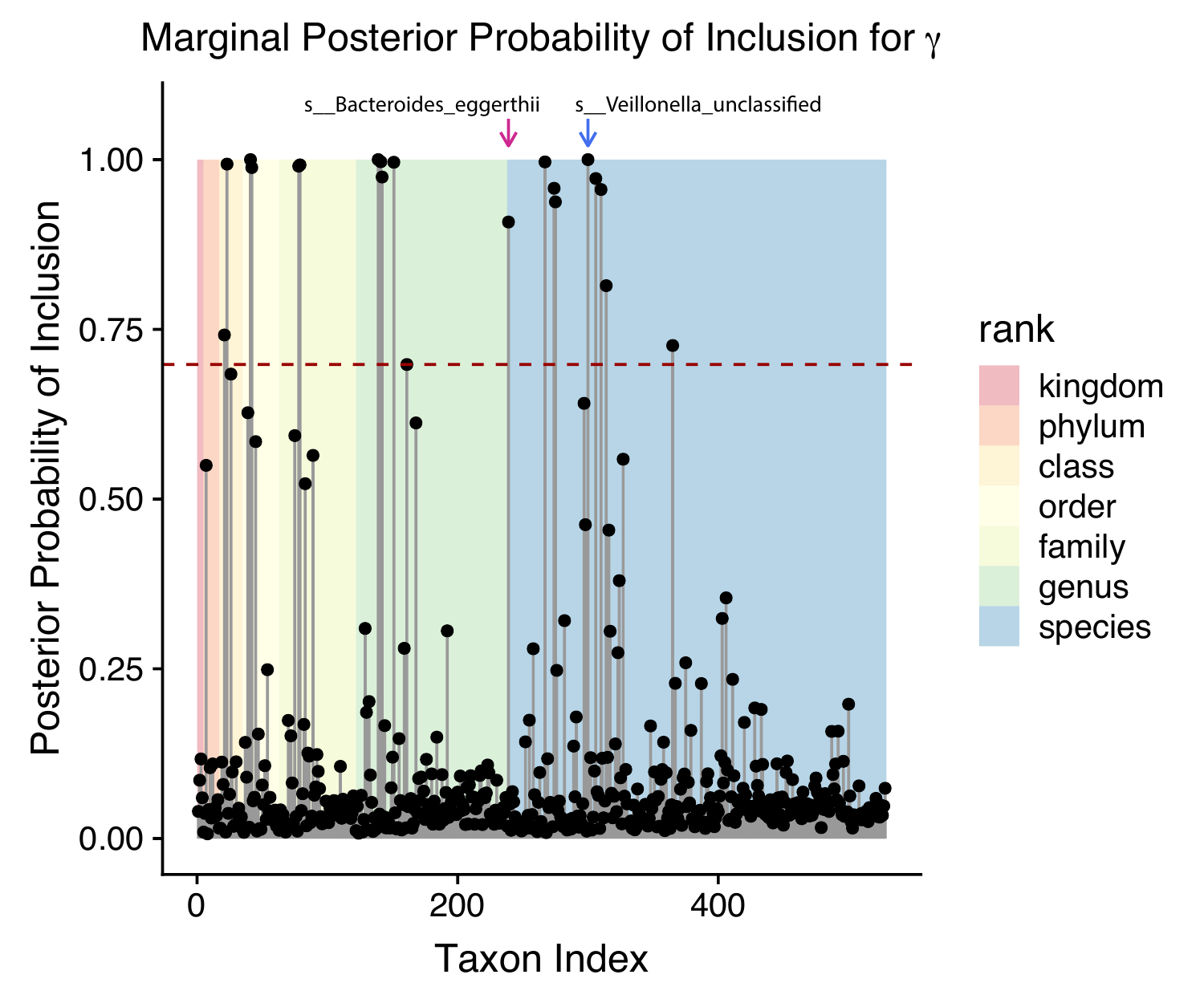}\label{realPPI1}} & 
		\subfloat[][Metastatic melanoma study: plot of $\gamma$ PPI]{\includegraphics[width=.5\textwidth]{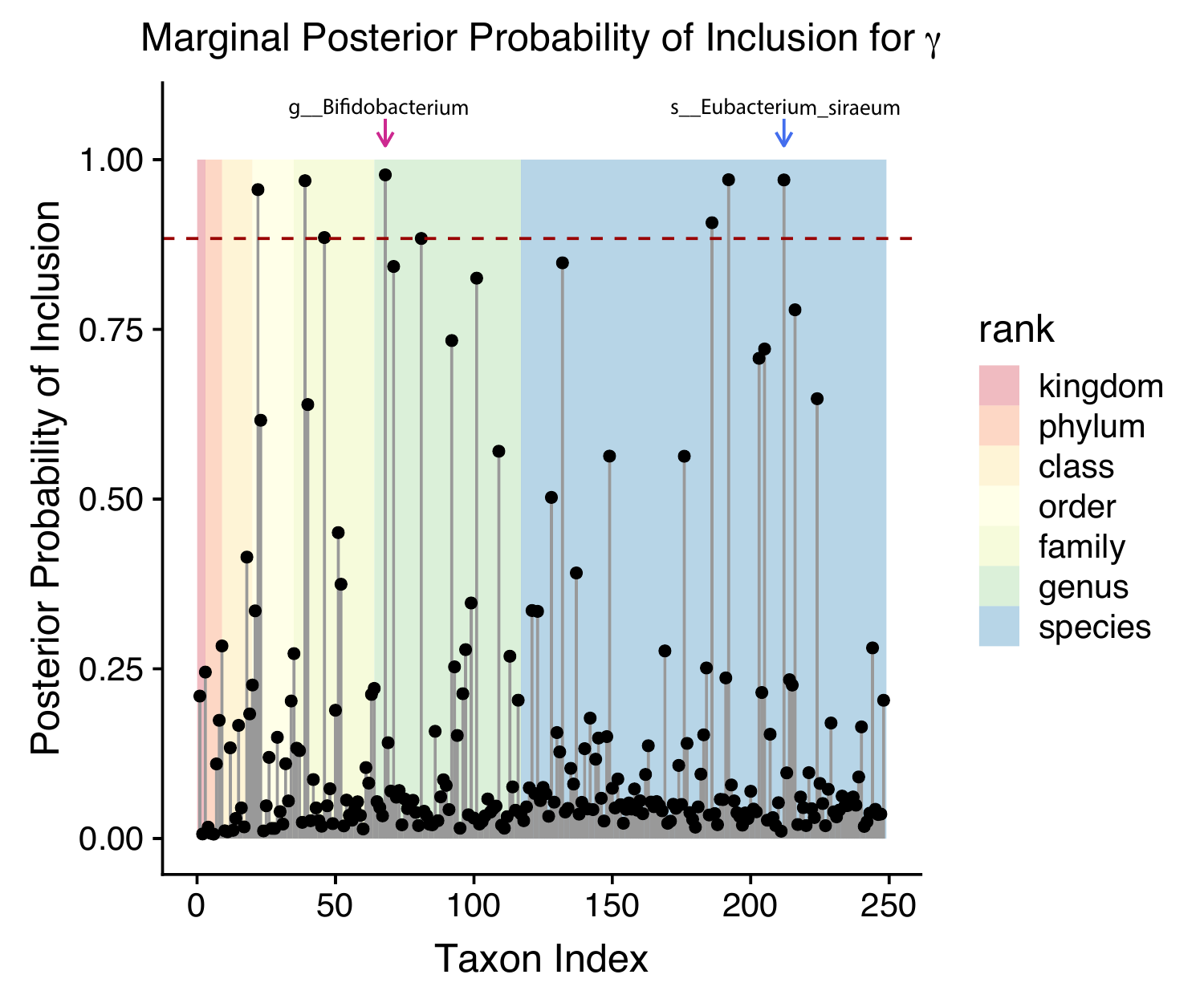}\label{realPPI2}}\\
		\subfloat[][Liver cirrhosis study: credible interval for $\mu_{2j}$]{\includegraphics[width=.5\textwidth]{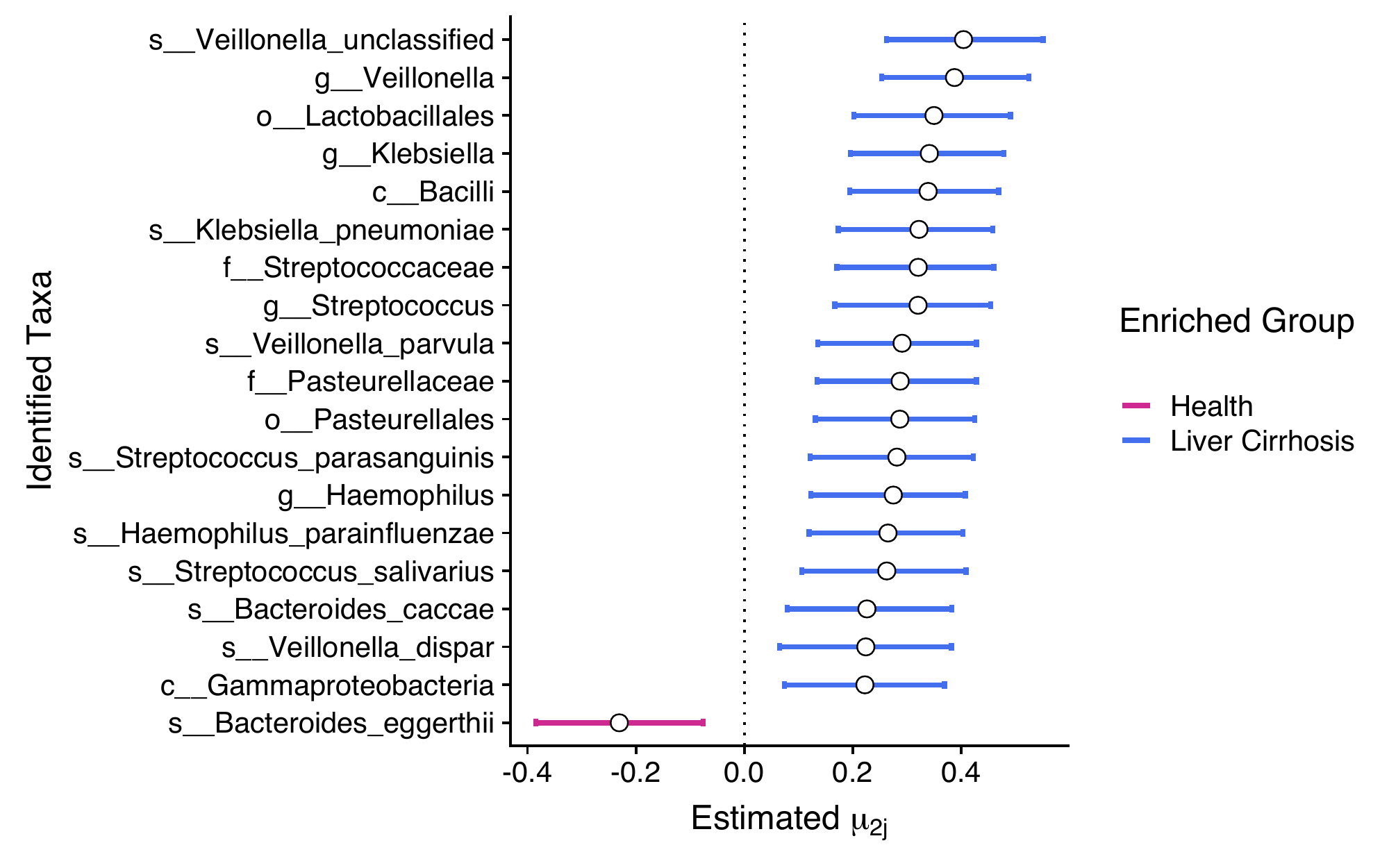}\label{CI1}} &
		\subfloat[][Metastatic melanoma study: credible interval for $\mu_{2j}$ ]{\includegraphics[width=.5\textwidth]{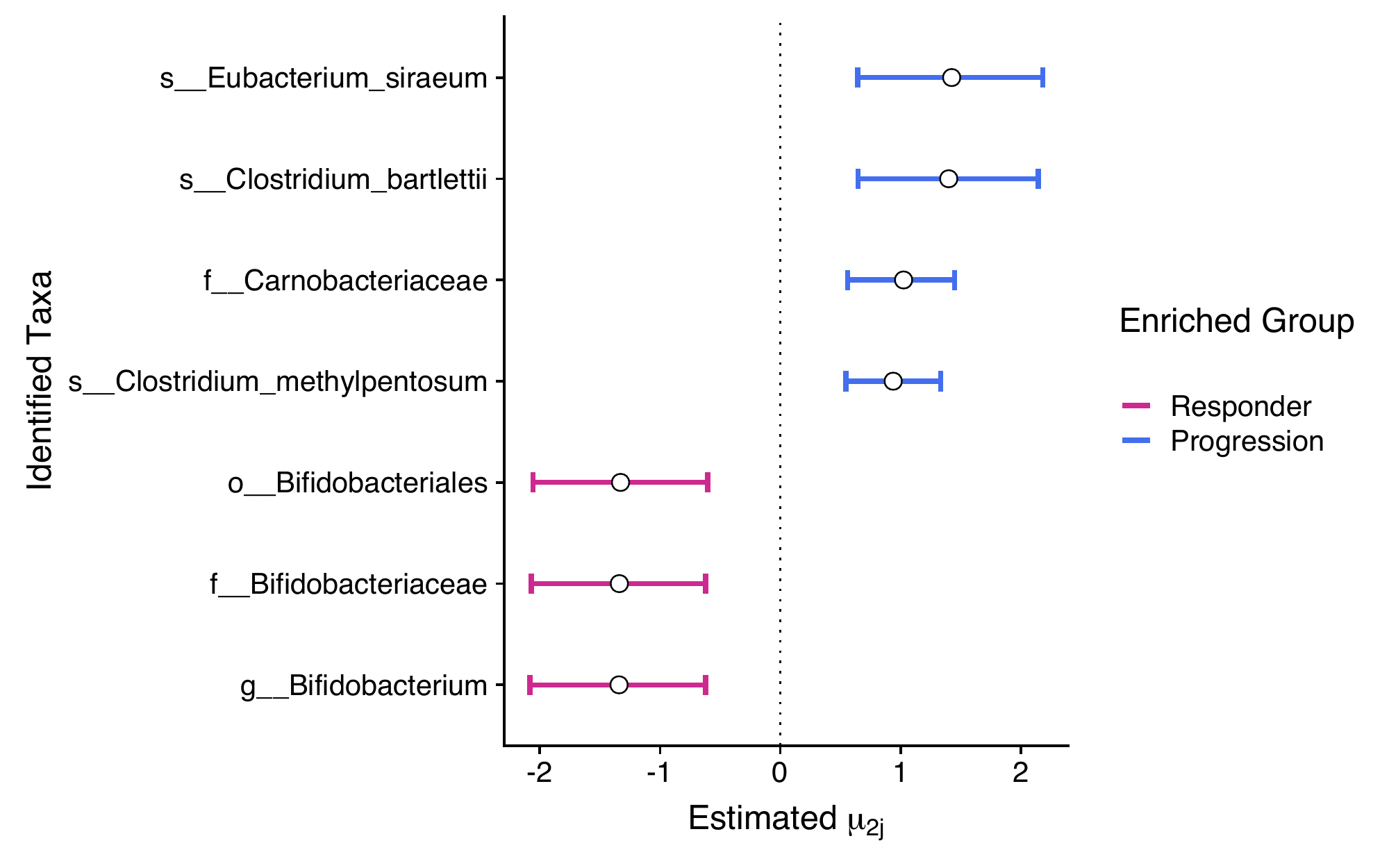}\label{CI2}}\\
	\end{tabular}
	\caption[]{Real Data Analysis: Plots for $\gamma$ PPI and credible interval. The horizontal dashed line in the PPI plot represents the threshold controlling the Bayesian false discovery rate $<$ 0.05. All taxa whose PPI pass the threshold are included in (c) and (d), where each horizontal bar is the 95\% credible interval for $\mu_{2j}$ (group-specific parameter) with posterior mean shown in circle. Each arrow in (a), (b) points out the taxon with largest absolute value of $\mu_{2j}$ in one patient group as shown in Figure \ref{CI1} and \ref{CI2}.}
	\label{fig:realdata}
\end{figure}

\begin{figure}
	\centering
	\subfloat[Liver cirrhosis study: Cladogram for discriminating taxa]{%
		\includegraphics[width=0.8\columnwidth]{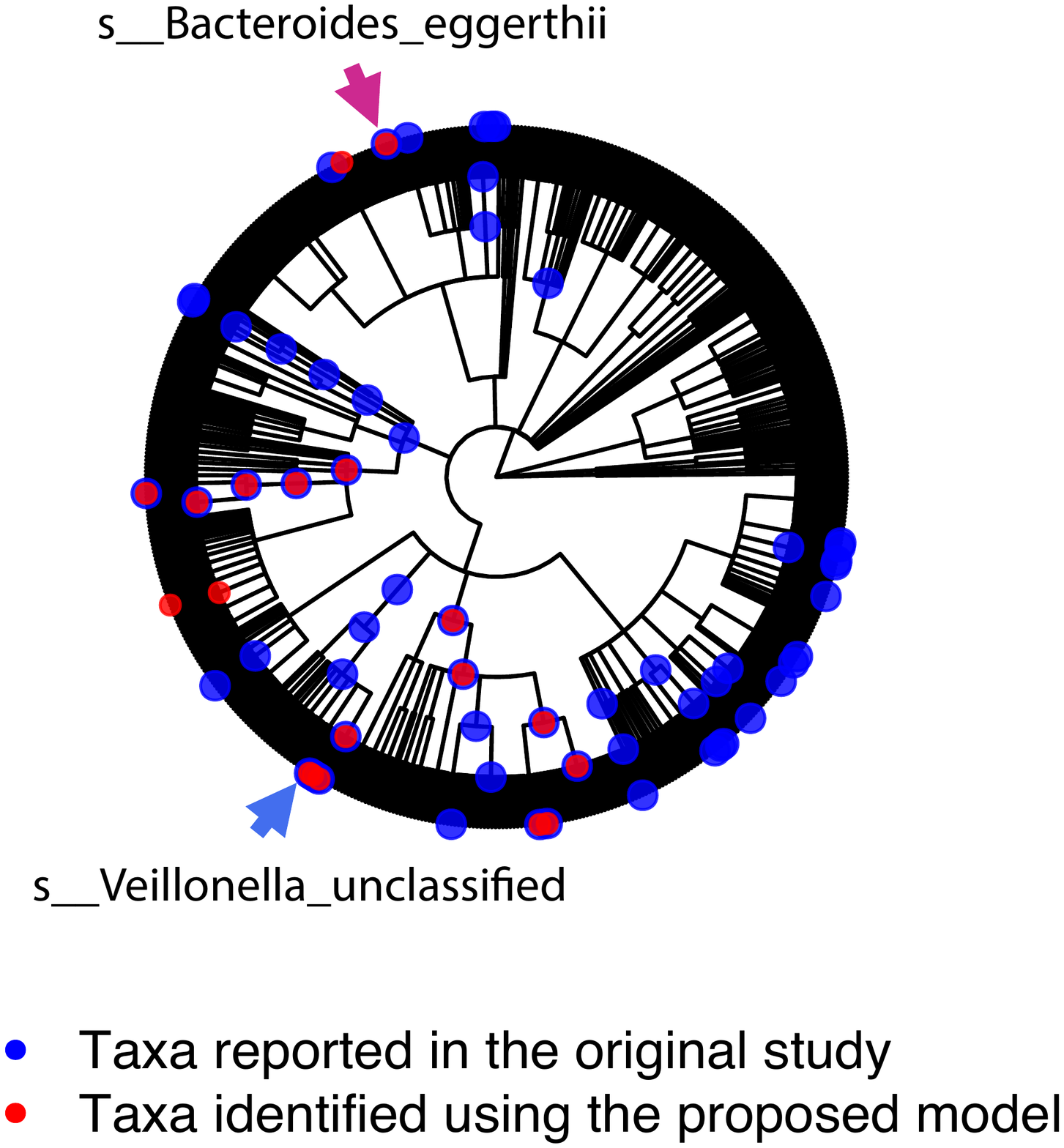}\label{cladogram1}%
	}
	
	\subfloat[Metastatic melanoma study: Cladogram for discriminating taxa]{%
		\includegraphics[width=0.8\columnwidth]{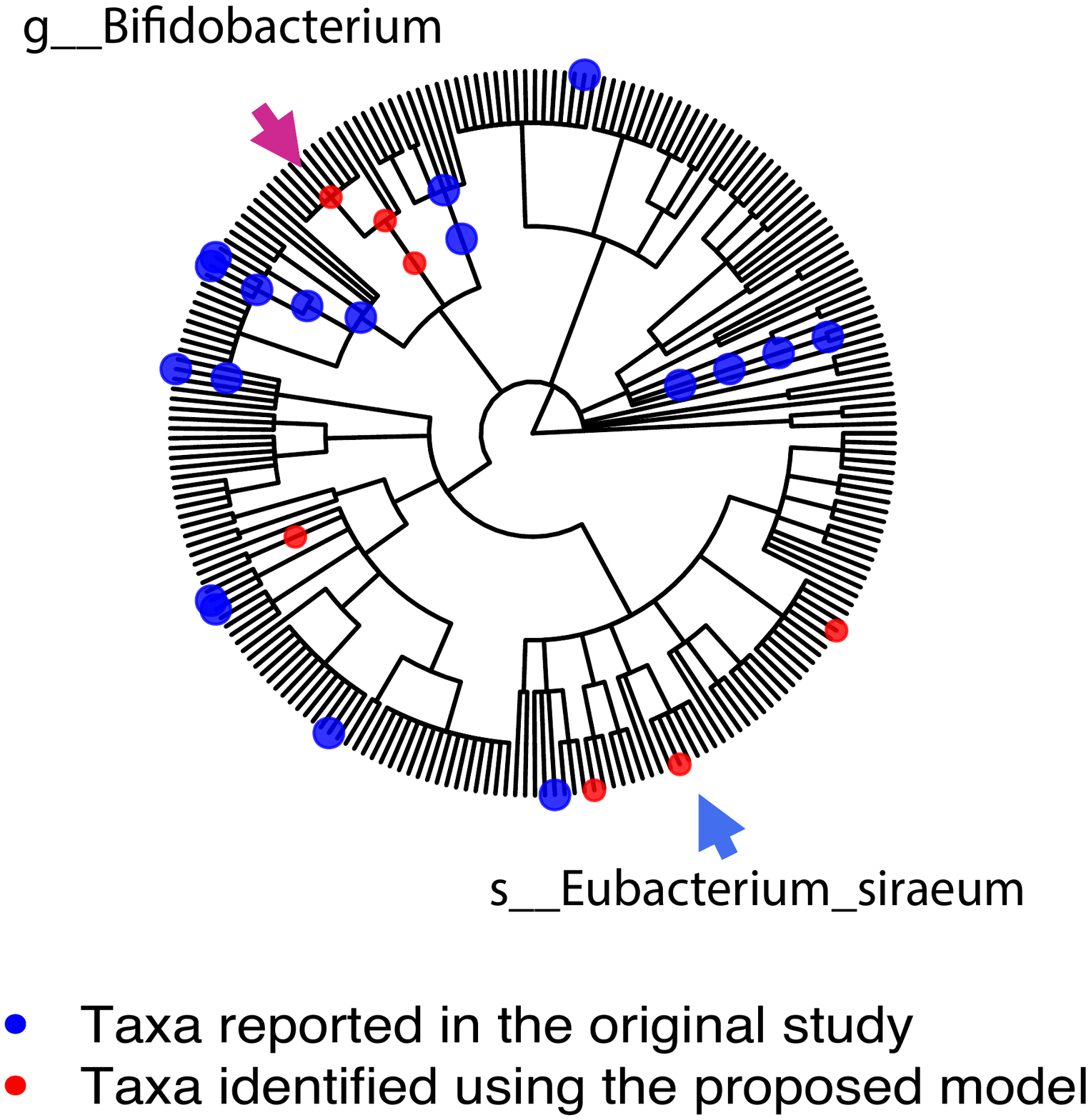}\label{cladogram2}%
	}
	
	\caption[]{Real Data Analysis: Cladograms of the identified discriminating taxa (shown in dots). Red dots: taxa found by the proposed model; Blue dots: taxa found by methods reported in the original studies. Each arrow in (a), (b) points out the taxon with the largest absolute value of $\mu_{2j}$ (group-specific parameter) in one patient group, as shown in Figure \ref{CI1} and \ref{CI2}.}
	
	\label{fig:cladogram}
\end{figure}

In the model fitting stage, for prior specification, we used $a_p = 0.2, ~b_p = 1.8$ to obtain a sparser covariate effect due to the small sample size, and it suggested about 10\% of taxa-covariate associations were significant. We kept the same default setting for the rest of the  hyperparameters. Next, we ran four independent chains with different starting points, and discarded the first half of 40,000 iterations for each chain. Although the small sample size ($n = 24$) posed challenges for parameter estimation, the results showed high pairwise Pearson correlations of PPIs for $\gamma$ (ranging from 0.989 to 0.992) and $\delta$ (ranging from 0.927 to 0.953). Figure \ref{realPPI2} shows the PPIs for all taxa, and Figure 4d illustrates the posterior means of the selected taxa. Table S2 in the supplement includes detailed parameter estimations of the taxa in Figure 4d.

Our model jointly identified differentially abundant taxa and revealed the microbiome-metabolite associations. First, among all seven taxa identified, it is of specific interest to investigate the responder-enriched taxon \textit{Bifidobacterium} (genus level), \textit{Bifidobacteriaceae} (family level). \textit{Bifidobacterium}, nesting within \textit{Bifidobacteriaceae}, is a genus of gram-positive, nonmotile, often branched anaerobic bacteria \citep{schell2002genome}. \textit{Bifidobacteria} are one of the major genera of bacteria that make up the gastrointestinal tract microbiota in mammals. This result about \textit{Bifidobacterium} is supported by recent melanoma studies. \cite{sivan2015commensal} compared melanoma growth in mice harboring specific microbiota, and used sequencing of the 16S ribosomal RNA to identify \textit{Bifidobacterium} as associated with the antitumor effects. They also found that oral administration of \textit{Bifidobacterium} augmented ICT efficacy. Moreover, \cite{matson2018commensal} detected significant association between several species from \textit{Bifidobacterium} with patients' outcomes in an immunotherapy treatment study for metastatic melanoma. Both studies showed consistent direction of effect, as did our model. The responder-enriched taxon \textit{Bifidobacterium} were estimated to negatively correlate with 2-oxoarginine and 2-hydroxypalmitate in Figure \ref{fig:realdata}d. The suppression of these fatty-acid metabolites may induce better cancer treatment as they were shown to have the oncogenic signaling role in cancer cells \citep{louie2013cancer}.  


\section{Conclusion}\label{conclusion}
In this paper, we presented a Bayesian ZINB model for analysis of high-throughput sequencing microbiome data. Our method is novel in simultaneously incorporating the effect from measurable genetic covariates and identifying differentially abundant taxa for multiple patient groups in one statistical framework. This allows for integrative analysis of microbiome data and other omics data. Our method is flexible, as it allows for identification and estimation of the association between covariates and each taxon's abundance. These results could potentially guide clinical decisions for precision shaping of the microbiome, although results would need to be validated in preclinical models first. In addition, our method is computationally efficient in posterior inferences. We implemented the MCMC algorithm to analyze the data from two MSS studies with results readily available in minutes. 

In real data analysis, the identified differentially abundant taxa by our model are often cluttered in the same phylogenetic branch. These results are achieved without imposing the phylogenetic structures in the model. This highlights that the results from our model are biologically interpretable and thus capable of guiding further biological mechanism studies. Our results on the metastatic melanoma study uncover novel relationships between taxa and metabolites which merit further experimental investigation.

The framework of our model allows for several extensions. For example, the current method supports two phenotype groups. If there are multiple groups (e.g. the intermediate phenotypes), the current model can incorporate group-specific parameters while holding the other parameters unchanged (e.g. the \bch normalized \ech microbiome abundance can be inferred in the same way). Then the same posterior inferences can be applied. \bch The proposed model based on a regression framework considers the microbiome normalized abundance as the response and integrates the omics data, e.g. metabolite compounds, as predictors. Similarly, \cite{lloyd2019multi} used the microbial abundance as the response and a type of genomics data (i.e. single nucleotide polymorphism, SNP) as predictors to identify several inflammatory bowel disease (IBD)-associated host-microbial interactions. Both methods focus on the omics effect on microbial abundance. However, the interaction between the microbiome and the host is bidirectional. Therefore, it is worthwhile to consider using the microbial features as predictors to investigate their modulations on any biological process with quantitative omics measurements. For instance, \cite{richards2019gut} explored how microbial abundances induced changes in chromatin accessibility and transcription factor binding of host genetics. \ech Another interesting extension would be to analyze correlated covariates such as longitudinal clinical measurements \citep{zhang2017negative}. 

\section{Software}
\bch An \verb|R| package \verb|IntegrativeBayes| is available on GitHub:
 
\noindent\url{https://github.com/shuangj00/IntegrativeBayes}.\ech  All the simulated datasets and two real datasets presented in Section \ref{realdata} are available on figshare: 
 
\noindent\url{https://figshare.com/projects/IntegrativeBayesZINB/57980}.

\section{Supplementary Material}
\label{sec6}

Supplementary material is available online at
\url{http://biostatistics.oxfordjournals.org}.

\section*{Acknowledgments}

The authors thank Jiwoong Kim for processing the microbiome data and initiating helpful discussions. We thank Jessie Norris for her comments on the manuscript. \bch We thank the authors of \cite{li2018conditional} for sharing the \texttt{R} codes to evaluate the MZILN model. \ech We thank the editor and two reviewers for their comments that led to improvements in the manuscript. 
{\it Conflict of Interest}: Dr. Andrew Koh is a consultant for Merck and Saol Therapeutics.

\bibliographystyle{biorefs}
\bibliography{biostatistics_manuscript_0927}

\clearpage

\textbf{
	Supplementary Material}

\maketitle

\section*{S1. Details of MCMC Algorithms}
First, we write the likelihood function as follows:
\begin{align*}
&\prod_{k=1}^K \prod_{i:z_i = k}  \prod_{j: \gamma_j = 1, r_{ij}= 0} \frac{\Gamma(y_{ij}+\phi_j)}{y_{ij!}\Gamma(\phi_j)}\left ( \frac{\phi_j}{s_i e^{\mu_{0j}+\mu_{kj}+\bm{x}_i\boldsymbol{\beta}_j^T}+\phi_j} \right )^{\phi_j}\left ( \frac{s_i e^{\mu_{0j}+\mu_{kj}+\bm{x}_i\boldsymbol{\beta}_j^T}}{s_i e^{\mu_{0j}+\mu_{kj}+\bm{x}_i\boldsymbol{\beta}_j^T}+\phi_j}  \right )^{y_{ij}} \times \\ 
&\prod_{i} \prod_{j: \gamma_j = 0, r_{ij}= 0} \frac{\Gamma(y_{ij}+\phi_j)}{y_{ij!}\Gamma(\phi_j)}\left ( \frac{\phi_j}{s_ie^{\mu_{0j}+\bm{x}_i\boldsymbol{\beta}_j^T}+\phi_j} \right )^{\phi_j}\left ( \frac{s_ie^{\mu_{0j}+\bm{x}_i\boldsymbol{\beta}_j^T}}{s_ie^{\mu_{0j}+\bm{x}_i\boldsymbol{\beta}_j^T}+\phi_j}  \right )^{y_{ij}} .
\end{align*}
Then, we update the parameters in each iteration following the steps below:
\begin{enumerate}
	\item \textbf{Update of zero-inflation latent indicator} $r_{ij}$: Notice that we only need to update the $r_{ij}$'s that correspond to $y_{ij} = 0$. 
	We write the posterior as:
	\begin{align*}
	&p(r_{ij}|y_{ij}=0,\phi_j,z_i = k,s_i,\mu_{0j},\mu_{kj},\gamma_j)  \\ 
	\propto & \int L(r_{ij}|y_{ij}=0,\phi_j,z_i = k,\mu_{0j},\mu_{kj},\gamma_j) \times p(r_{ij}|\pi ) \times p(\pi) d\pi 
	\end{align*}
	Then it follows that
	\begin{align*}
	p(r_{ij}|\cdot) \propto \left\{\begin{matrix}
	\left ( \frac{\phi_j}{s_i e^{\mu_{0j}+\mu_{kj}+\bm{x}_i\boldsymbol{\beta}_j^T}+\phi_j} \right )^{\phi_j(1-r_{ij})} \times \frac{Be(a_{\pi}+r_{ij},b_{\pi}-r_{ij}+1)}{Be(a_{\pi}, b_{\pi})} & if~\gamma_j = 1\\\ 
	\left ( \frac{\phi_j}{s_i e^{\mu_{0j}+\bm{x}_i\boldsymbol{\beta}_j^T}+\phi_j} \right )^{\phi_j(1-r_{ij})} \times \frac{Be(a_{\pi}+r_{ij},b_{\pi}-r_{ij}+1)}{Be(a_{\pi}, b_{\pi})}& if~\gamma_j = 0
	\end{matrix}\right.
	\end{align*} 
	\item \textbf{Update of} $\boldsymbol{\mu}_0$: We update each $\mu_{0j},~j=1,2,...p$ sequentially using an independent Metropolis-Hasting algorithm. We first propose a new $\mu_{0j}^*$ from $N(\mu_{0j},\tau_0^2)$ and then accept the proposed value with probability min($1,~m_{MH})$, where
	\begin{align*}
	m_{MH} = \frac{\prod_{i=1}^nf(y_{ij}|\mu_{0j}^*,\mu_{kj},\phi_j,s_i,\gamma_j,\bm{R},\bm{X},\bm{B}) \times p(\mu_{0j}^*) \times J(\mu_{0j};\mu_{0j}^*)}{\prod_{i=1}^nf(y_{ij}|\mu_{0j},\mu_{kj},\phi_j,s_i,\gamma_j,\bm{R},\bm{X},\bm{B}) \times p(\mu_{0j}) \times J(\mu_{0j}^*;\mu_{0j})}
	\end{align*}
	\item \textbf{Joint Update of $\mu_{k\cdot}~\text{and}~\boldsymbol{\gamma}$}: A between-model step is implemented first to jointly update $\mu_{k\cdot}$ and $\boldsymbol{\gamma}$. We use an \textit{add-delete} algorithm, where we select a $j \in \{1, \ldots, p\}$ at random and change the value of $\gamma_j$. For the \textit{add} case, i.e. $ \gamma_j = 0 \rightarrow \gamma_j=1$, we propose $\mu^*_{kj}$ for each $k = 2,\ldots ,n$ from $N(0,\tau^2_{\mu j})$. For the \textit{delete} case, i.e. $ \gamma_j = 1 \rightarrow \gamma_j=0$, we set $\mu^*_{kj}=0$ for all $k$. We finally accept the proposed values with probability min($1,~m_{MH}$), where
	\begin{align*}
	m_{MH} = &\frac{\prod_{i=1}^nf(y_{ij}|\mu_{kj}^*,\mu_{0j},\phi_j,s_i,\gamma_j^*,\bm{R},\bm{X},\bm{B}) \times p(\mu_{k j}^*|\gamma_j^*) \times p(\boldsymbol{\gamma}^*) }{\prod_{i=1}^nf(y_{ij}|\mu_{kj},\mu_{0j},\phi_j,s_i,\gamma_j,\bm{R},\bm{X},\bm{B}) \times p(\mu_{k j}|\gamma_j) \times p(\boldsymbol{\gamma})} \\
	&\times  \frac{J(\mu_{k j};\mu_{k j}^*|\gamma_j;\gamma_j^*) \times J(\boldsymbol{\gamma};\boldsymbol{\gamma}^*)}{ J(\mu_{k j}^*;\mu_{k j}|\gamma_j^*;\gamma_j) \times J(\boldsymbol{\gamma}^*;\boldsymbol{\gamma})}
	\end{align*} 
	
	\textbf{Further update of $\mu_{kj}$ when $\gamma_j ^*= 1$}: A within-model step is followed to further update each $\mu_{kj},~k=2,\ldots, K$ that corresponds to $\gamma_j^*=1$ in the current iteration. We first propose a new $\mu_{kj}^*$ from $N(\mu_{kj},(\tau_{\mu j}/2)^2)$ and then accept the proposed value with probability min($1,m_{MH}$), where
	\begin{align*}
	m_{MH} = \frac{\prod_{i=1}^nf(y_{ij}|\mu_{kj}^*,\mu_{0j},\phi_j, s_i,\gamma_j,\bm{R},\bm{X},\bm{B}) \times p(\mu_{kj}^*) \times J(\mu_{kj};\mu_{kj}^*)}{\prod_{i=1}^nf(y_{ij}|\mu_{kj},\mu_{0j},\phi_j, s_i,\gamma_j,\bm{R},\bm{X},\bm{B}) \times p(\mu_{kj}) \times J(\mu_{kj}^*;\mu_{kj})}
	\end{align*} 
	
	\item \textbf{Joint update of $\beta_{\cdot j}$ and $\delta_{\cdot j}$}: Very similar to the above, we perform a between-model step first using an \textit{add-delete} algorithm. For each $j =1,\ldots, p$, we first select an $r \in  \{1, \ldots, R\}$ at random and change the value of $\delta_{rj}$. For the \textit{add} case, i.e. $ \delta_{rj} = 0 \rightarrow \delta_{rj} = 1$, we propose $\beta^*_{rj}$ from $N(0,\tau^2_{\beta j})$. For the \textit{delete} case, i.e. $ \delta_{rj} = 1 \rightarrow \delta_{rj} = 0$, we set $\beta^*_{rj}=0$. Then finally we accept the proposed values with probability min($1,m_{MH}$), where
	\begin{align*}
	m_{MH} = &\frac{\prod_{i=1}^nf(y_{ij}|\beta_{rj}^*,\delta_{rj}^*, \mu_{0j},\mu_{\cdot j},s_i,\gamma_j,\bm{R},\bm{X})  \times p(\beta_{\cdot j}^*|\delta_{\cdot j}^*) \times p(\delta_{\cdot j}^*)}{\prod_{i=1}^nf(y_{ij}|\beta_{rj},\delta_{rj},\mu_{0j},\mu_{\cdot j},s_i,\gamma_j,\bm{R},\bm{X}) \times p(\beta_{\cdot j}|\delta_{\cdot j}) \times p(\delta_{\cdot j})} \\
	&\times \frac{J(\beta_{\cdot j};\beta_{\cdot j}^*|\delta_{\cdot j};\delta_{\cdot j}^*) \times J(\delta_{\cdot j};\delta_{\cdot j}^*)}{ J(\beta_{\cdot j}^*;\beta_{\cdot j}|\delta_{\cdot j}^*;\delta_{\cdot j}) \times J(\delta_{\cdot j}^*;\delta_{\cdot j})}
	\end{align*} 
	
	\textbf{Further update of $\beta_{rj}$ when $\delta_{rj}^* = 1$}: A within-model step is followed to further update each $\beta_{rj},~r=1,\ldots,R$ that corresponds to $\delta_{rj}^*=1$. We first propose a new $\beta_{rj}^*$ from $N(\beta_{rj},(\sigma_{\beta j}/2)^2)$ and then accept the proposed value with probability min($1,m_{MH}$), where
	\begin{align*}
	m_{MH} =\frac{\prod_{i=1}^nf(y_{ij}|\beta_{rj}^*,\delta_{rj},\mu_{0j},\mu_{\cdot j},s_i,\phi_j,\gamma_j,\bm{R},\bm{X}) \times p(\beta_{rj}^*) \times J(\beta_{rj};\beta_{rj}^*)}{\prod_{i=1}^nf(y_{ij}|\beta_{rj},\delta_{rj},\mu_{0j},\mu_{\cdot j},s_i,\phi_j,\gamma_j,\bm{R},\bm{X}) \times p(\beta_{rj}) \times J(\beta_{rj}^*;\beta_{rj})}
	\end{align*}
	\item \textbf{Update of $\phi_{\cdot}$}: We update each $\phi_j$ $j= 1, \ldots, p$ sequentially by using an independent Metropolis-Hasting algorithm. We first propose a new $\phi_j^*$ from the normal distribution $N(\phi_j, \tau_{\phi}^2)$ that truncated at 0, and accept the proposed value with probability min(1,$m_{MH}$), where
	\begin{align*}
	m_{MH} = \frac{\prod_{i=1}^nf(y_{ij}|\phi_j^*,\beta_{rj},\delta_{rj},\mu_{0j},\mu_{\cdot j},s_i,\gamma_j,\bm{R},\bm{X},\bm{B}) \times p(\phi_{j}^*) \times J(\phi_{j};\phi_{j}^*)}{\prod_{i=1}^nf(y_{ij}|\phi_{j},\beta_{rj},\delta_{rj},\mu_{0j},\mu_{\cdot j},s_i,\gamma_j,\bm{R},\bm{X}, \bm{B}) \times p(\phi_{j}) \times J(\phi_{j}^*;\phi_{j})}
	\end{align*}
\end{enumerate}

\section*{S2. Results of Simulation Study}\label{simulation_supp}
We performed a comprehensive simulation study for model comparison. First, we introduce the following reference setting in the simulation study, that is, 1) $n = 60$ samples split into $K=2$ equally sized groups; 2) $p=300$ features, $20$ of which were truly discriminating ones; 3) $\pi_0 = 40\%$ false zeros (i.e. structural zeros) randomly assigned among all counts; 4) $R=7$ covariates, four of which true coefficients were nonzero; 5) noise level $\epsilon_e^2 = 1$. Furthermore, we varied the following settings to comprehensively examine the model performance, including the choices for sample size per group ($n / 2=10$ or $30$), the three log-scale noise levels ($\sigma_e=0.5$, $1.0$, or $1.5$) and the extra zero proportions ($\pi_0 =30\%$, $40\%$, or \bch $70\%$\ech). In all cases, we randomly set four out of seven nonzero $\beta_{rj}$ for each taxon $j$. 

\subsection*{S2.1 Evaluation for Sample Size}
Figure \ref{sample_supp} compares the model performance under two choices of group sizes ($n / 2=10$ or $30$) with fixed log-scale noise level at $1.0$ and $40\%$ of extra zeros as in the reference setting. Different methods are compared using the receiving operating characteristic (ROC) curve and area under the curve (AUC). The left part in Figure \ref{sample_supp} shows the results of identifying the differentially abundant taxa ($\bm{\gamma})$ and the right part is the results of detecting significant covariate-taxa associations ($\bm{\Delta}$). Clearly, decreasing the sample size hampers the performance of all the methods, but the proposed ZINB model maintains the highest AUC in both cases, and achieves the highest true positive rate under a fixed small false positive rate.

\subsection*{S2.2 Evaluation for Log-scale Noise Level}
Figure \ref{noise_supp} compares the model performance under three choices of log-scale noise level ($\sigma_e=0.5$, $1.0$, or $1.5$) with fixed group size of $n/2  =30$ and $40\%$ extra zeros as in the reference setting. The ZINB model maintains the highest AUC across all settings of identifying the differentially abundant taxa (AUC $> 0.9$), and detecting significant covariate-taxa associations (AUC $>0.8$). Notice that the true log-scale signal level is set to be $2$, and the ZINB model still shows an obvious advantage over the alternative methods under a large log-scale noise level of $1.5$.

\subsection*{S2.3 Evaluation for Extra Zero Proportion}
\bch
Figure S5 compares the model performance under three scenarios of extra zero proportions ($\pi_0=30\%$, $40\%$, and $70\%$) with a group size of $n/2 = 30$ and a noise level of $\epsilon_e^2=1$ as in the reference setting. Although a higher proportion of zeros like $70\%$ dose downgrade the performance of all methods, the proposed ZINB model is the best under any circumstance. Particularly, our methods always exhibits a considerably advantage over the others in terms of the identification of feature-covariate association ($\bm{\Delta}$) as shown in the right column of Figure S5.
\ech


\section*{S3. Sensitivity Analysis}\label{sensitivity_all}

\bch
\subsection{Choice of Size Factor Estimation}\label{size_factor_supp}
To assess model robustness with respect to the choice of size factor estimation methods, we compared the model performance under five typical normalization methods for the analysis of high-dimensional count data. They are: 1) geometric mean of pairwise ratios (GMPR) proposed by \cite{chen2018gmpr}; 2) cumulative sum scaling (CSS) proposed by \cite{Paulson2013}; 3) The 0.75-th quantile (Q75) proposed by \cite{Bullard2010}; 4) trimmed mean of M values (TMM) proposed by \cite{Robinson2010}; 5) relative log expression (RLE) proposed by \cite{Anders2010}. The first two, designed for normalizing the microbiome count data, have been described in Section 2.3 in the main text, while Q75, TMM and RLE are commonly used in RNA-seq data studies. In particular, Q75 calculates the size factor based on the upper-quantile (75\%) of the count distribution of a sample. TMM first sets a reference sample, and calculates the trimmed mean of the log ratios between all other samples with the selected reference to estimate size factors. RLE, on the other hand, computes a reference value of each feature (taxon) as the geometric mean across all samples, and then obtains ratios by dividing all features by the reference. The size factor for a sample by RLE is set to be the median of the ratios. Due to the high sparsity observed in the microbiome data, it is needed to add a pseudo-count such as 1 to the count matrix when using RLE to estimate size factors.

To test if the performance of our model is robust to the choice of different normalization techniques, we used the simulate datasets generated by the reference setting described in Section \ref{simulation}. The resulting AUCs for the discriminating feature indicator $\bm{\gamma}$ and the feature-covariate association indicator $\bm{\Delta}$ over $100$ data replicates are summarized in Figure S6. First, the result suggests that the proposed ZINB model is robust with respect to plug-in size factor estimations. Next, the ZINB-CSS and ZINB-GMPR show better performances due to the smaller variation and slightly higher average AUCs, since both are based on the normalization methods that better account for the characteristics of the microbiome data. Notice that RLE is less stable compared to the other methods for such sparse count data, which is also mentioned in \cite{chen2018gmpr}.

\ech

\subsection*{S3.2 Choice of Inverse-gamma Hyperparameters}\label{sensitivity}
We assess impacts of setting priors via sensitivity analysis. In our model, the choice of $a$ and $b$ in the IG$(a, b)$ prior for $\sigma_{\mu j}^2$ has an impact on the posterior probabilities of inclusion of $\gamma$. To investigate model performance with respect to the choice of these hyperparameters, we simulated 30 datasets under the reference setting described in Section \ref{simulation}, and benchmarked our model with varying values of $a$ from $0.5$ to $6$ and $b$ from $0.5$ to $25$. The choices of $a$ and $b$ are illustrated in Figure \ref{sensi}. 

The results given by different values of ($a,b$) were compared based on the Matthews correlation coefficient (MCC) \citep{matthews1985homeostasis} across 30 replicated datasets. In each replicate, we controlled a $5\%$ Bayesian false discovery rate and selected discriminating features. We then calculated the number of true positive (TP), true negative (TN), false positive (FP) and false negative (FN) and MCC. Here MCC is defined as
\begin{align*} \text{MCC}=\frac{\text{TP}\times\text{TN}-\text{FP}\times\text{FN}}{\sqrt{(\text{TP}+\text{FP})(\text{TP}+\text{FN})(\text{TN}+\text{FP})(\text{TN}+\text{FN})}}~.
\end{align*}
MCC ranges from $-1$ to $1$, and larger values represents favorable prediction results. It is also demonstrated in the above formula that the MCC-based evaluation is suitable for classes with very different sizes, since it strikes a balance between TP and FP counts. In our scenario, the size of truly discriminating features are relatively small compared to the total number. Therefore, we adopt MCC as an appropriate performance metric to handle the imbalanced setting. As can be seen in Figure \ref{sensi}, given a small value of $b$ ($b\leq 2$), the MCC is undesirable with any value of $a$ displayed here. As shown by \cite{gelman2006prior}, the IG($a,b$) prior with small $a$ and $b$ would distort the posterior inferences. On the other hand, if we increase $a$ to have $a> 2$ while fixing $b$ to be small, the corresponding prior distribution is strongly informative since IG($a,b$) has the mean of $b / (a - 2)$ and the variance of $b^2/(a- 2)^2(a - 1)$. Therefore, we choose $a = 2$ and $b = 10$ to be the default setting, since this ensures a flat prior and yields a beneficial variable selection result as shown in Figure \ref{sensi}.  


\begin{figure}[!h]
	\centering
	\includegraphics[width = \linewidth]{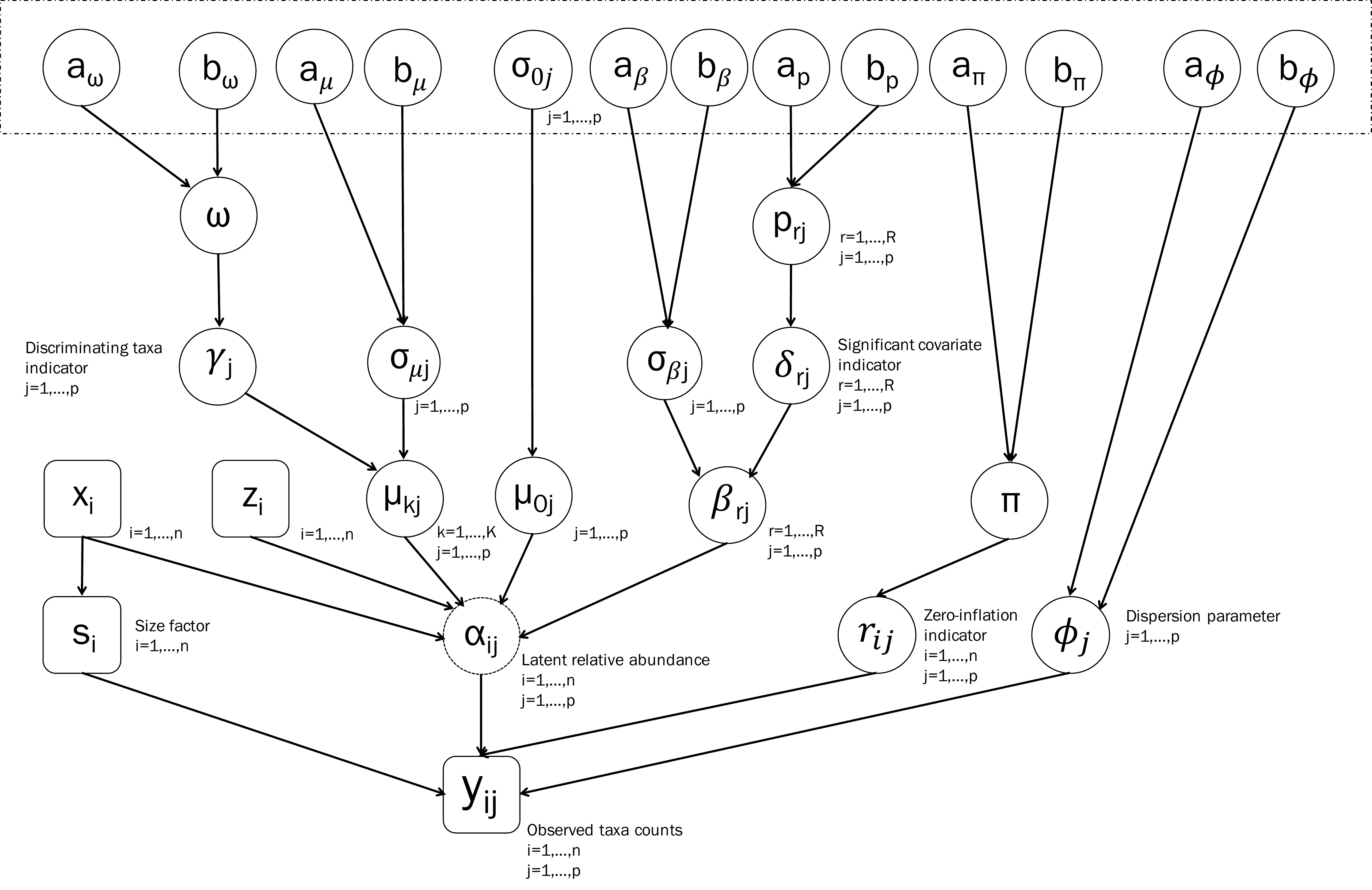}
	\caption{The graphical formulation of the proposed Bayesian zero-inflated negative binomial regression model. Node in a circle refers to a parameter of the model. Node in a rectangle is observable data. Circle nodes in the dashed block are fixed hyperparameters. The link between two nodes represents a direct probabilistic dependence.}
	\label{f1}
\end{figure}
\begin{figure}[!h]
	\hskip .1in
	\footnotesize
	\resizebox {0.99\textwidth }{!}{
		\begin{tabular}{|p{16cm}|}
			\hline
			{\textbf{Parameters}:}
			\begin{eqnarray*}
				\bm{R}&=& [\bm{r}_{\cdot 1},\ldots,\bm{r}_{\cdot j},\ldots,\bm{r}_{\cdot p}]\\
				\bm{\phi} &=& [\phi_1,\ldots,\phi_j,\ldots,\phi_p]\\
				\boldsymbol{\mu}_0 &=& [\mu_{01}, ,\ldots, \mu_{0j},\ldots,\mu_{0p}]\\
				\bm{M} &=& [\bm{\mu}_{\cdot 1},\ldots,\bm{\mu}_{\cdot j},\ldots,\bm{\mu}_{\cdot p}]\\
				\bm{B} &=& [\bm{\beta}_{\cdot 1},\ldots,\bm{\beta}_{\cdot j},\ldots,\bm{\beta}_{\cdot p}]\\
				\boldsymbol{\gamma} &=& [\gamma_1,\ldots,\gamma_j,\ldots,\gamma_p] \\
				\boldsymbol{\Delta} &=& [\bm{\delta}_{\cdot 1},\ldots,\bm{\delta}_{\cdot j},\ldots,\bm{\delta}_{\cdot p}]
			\end{eqnarray*}
			\\
			\hline
			{\textbf{Mixture model likelihood}:}
			\begin{eqnarray*}
				y_{ij}|r_{ij} = 0, z_i=k,\gamma_j=1  &\stackrel{ind}{\sim}& \NB(y_{ij}; s_i \alpha_{ijk},\phi_j)\quad
				\text{with } \log(\alpha_{ijk})=\mu_{0j}+\mu_{kj}+\bm{x}_{i\cdot}^T\bm{\beta}_{\cdot j}\\
				y_{ij}|r_{ij} = 0, \gamma_j=0 &\stackrel{ind}{\sim}& \NB(y_{ij};s_i \alpha_{ij0},\phi_j)\quad
				\text{with } \log(\alpha_{ij0})=\mu_{0j}+\bm{x}_{i\cdot}^T\bm{\beta}_{\cdot j} \\
				y_{ij}|r_{ij} = 1  &\equiv &  0
			\end{eqnarray*}
			\\
			\hline
			{\textbf{Zero-inflation prior}:}
			\begin{eqnarray*}
				r_{ij}|\pi  \sim \text{Bernoulli}(\pi),\quad \pi \sim \text{Beta}(a_\pi,b_\pi) \quad \Rightarrow \quad 
				r_{ij}|a_\pi,b_\pi \sim \text{Beta-Bernoulli}(r_{ij};a_\pi,b_\pi)
			\end{eqnarray*}
			\\
			\hline
			{\textbf{Feature selection prior}:}
			\begin{eqnarray*}
				\gamma_{j}|\omega  \sim \text{Bernoulli}(\omega),\quad \omega \sim \text{Beta}(a_\omega,b_\omega) \quad \Rightarrow \quad 
				\gamma_{j}|a_\omega,b_\omega \sim \text{Beta-Bernoulli}(\gamma_{j};a_\omega,b_\omega)
			\end{eqnarray*}
			\\
			\hline
			{\textbf{Dispersion prior}:}
			\begin{eqnarray*}
				\phi_{j}  \sim  \Ga(a_\phi,b_\phi)
			\end{eqnarray*}
			\\
			\hline
			{\textbf{Feature / Covariate characterization priors}:}
			\begin{eqnarray*}
				\mu_{0j}|\sigma_{0j}\sim\normal(0,\sigma_{0j}^2)&&  \\
				\mu_{kj}|\gamma_j , \sigma_{kj}\sim (1 - \gamma_j)\I(\mu_{kj} = 0) + \gamma_{j}\normal(0,\sigma_{\mu j}^2)&,&\quad\sigma_{\mu j}^2\sim\IG(a_\mu,b_\mu) \Rightarrow\\
				&& \mu_{kj}|\gamma_j \sim (1 - \gamma_j)\I(\mu_{kj} = 0)+\gamma_j t_{2a_\mu}(0,b_\mu/a_\mu)\\
				\beta_{rj}|\delta_{rj},\sigma_{\beta j}\sim(1-\delta_{rj})\I(\beta_{rj} = 0)+\delta_{rj}\normal(0,\sigma_{\beta j}^2)&,&\quad\sigma_{\beta j}^2\sim\IG(a_\beta,b_\beta) \Rightarrow\\
				&&  \beta_{rj}|\delta_{rj}\sim(1-\delta_{rj})\I(\beta_{rj} = 0)+\delta_{rj}t_{2a_\beta}(0,b_\beta/a_\beta)
			\end{eqnarray*}
			\\
			\hline
			{\textbf{Covariate selection prior}:}
			\begin{eqnarray*}
				\delta_{rj}|p_{rj}  \sim \text{Bernoulli}(p_{rj}),\quad p_{rj} \sim \text{Beta}(a_p,b_p) \quad \Rightarrow \quad 
				\delta_{rj}|a_p,b_p \sim \text{Beta-Bernoulli}(\delta_{rj};a_p,b_p)
			\end{eqnarray*}
			\\
			\hline
			{\textbf{Fixed hyperparameters}:}
			\[\sigma_{0j}, a_\pi,b_\pi,a_\omega,b_\omega, a_\phi,b_\phi,a_\mu,b_\mu,a_\beta,b_\beta\]
			\\
			\hline
			{\textbf{Posterior}:}
			\begin{eqnarray*}
				p(\bm{R}, \bm{\phi},\boldsymbol{\mu}_0,	\bm{M},\bm{B},\boldsymbol{\gamma},\boldsymbol{\Delta}| \bm{Y},\bm{X})&& \propto \\
				&& \hspace{-150pt} 
				\prod_{k=1}^K \prod_{i:z_i = k}  \prod_{\substack{j: \gamma_j = 1 \\ ~~~r_{ij}= 0}}\NB(y_{ij}; s_i \alpha_{ijk},\phi_j) \times 
				\prod_{i}  \prod_{\substack{j: \gamma_j = 0 \\ ~~~r_{ij}= 0}}\NB(y_{ij}; s_i \alpha_{ij0},\phi_j) \\
				&& \hspace{-150pt} 
				\times \prod_{i,j}\text{Beta-Bernoulli}(r_{ij};a_\pi,b_\pi) 
				\times  \prod_{j}\Ga(\phi_j; a_\phi,b_\phi) 
				\times \prod_{j} \normal(\mu_{0j};0,\sigma_{0j}^2) 
				\\
				&&  \hspace{-150pt} 
				\times \prod_{j, k} [ (1 - \gamma_j)\I(\mu_{kj} = 0)+\gamma_j t_{2a_\mu}(\mu_{kj};0,b_\mu/a_\mu) ] \times \prod_{j}\text{Beta-Bernoulli}(\gamma_{j};a_\omega,b_\omega) \\
				&&  \hspace{-150pt} 
				\times 
				\prod_{r,j} [(1-\delta_{rj})\I(\beta_{rj} = 0)+\delta_{rj}t_{2a_\beta}(\beta_{rj};0,b_\beta/a_\beta) ]\times
				\prod_{r,j}\text{Beta-Bernoulli}(\delta_{rj};a_p,b_p)
			\end{eqnarray*}
			\\
			\hline
		\end{tabular}
	}
	\caption{Hierarchical formulation of the proposed hierarchical mixture model}
	\label{f1.1}
\end{figure}

\newpage
\begin{figure}[!h]
	\centering
	\begin{tabular}{@{}c@{}}
		\subfloat[][ROC curves for $\bm{\gamma}$ (left) and $\bm{\Delta}$ (right) with a sample size per group of \bch$n/2 = 30$ \ech ]{\includegraphics[width=\textwidth]{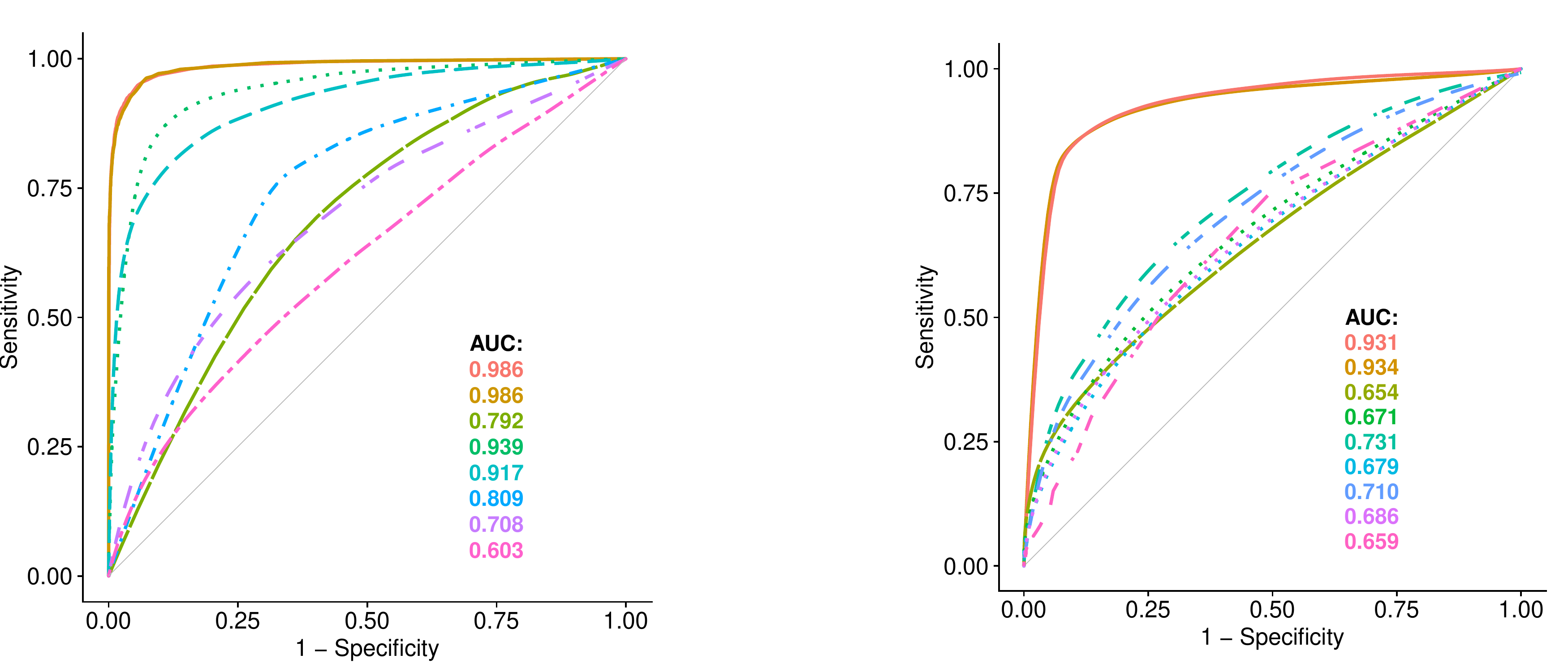}\label{size60}} \\
		\subfloat[][ROC curves for $\bm{\gamma}$ (left) and $\bm{\Delta}$ (right) with a sample size per group of \bch $n/2 = 10$ \ech ]{\includegraphics[width=\textwidth]{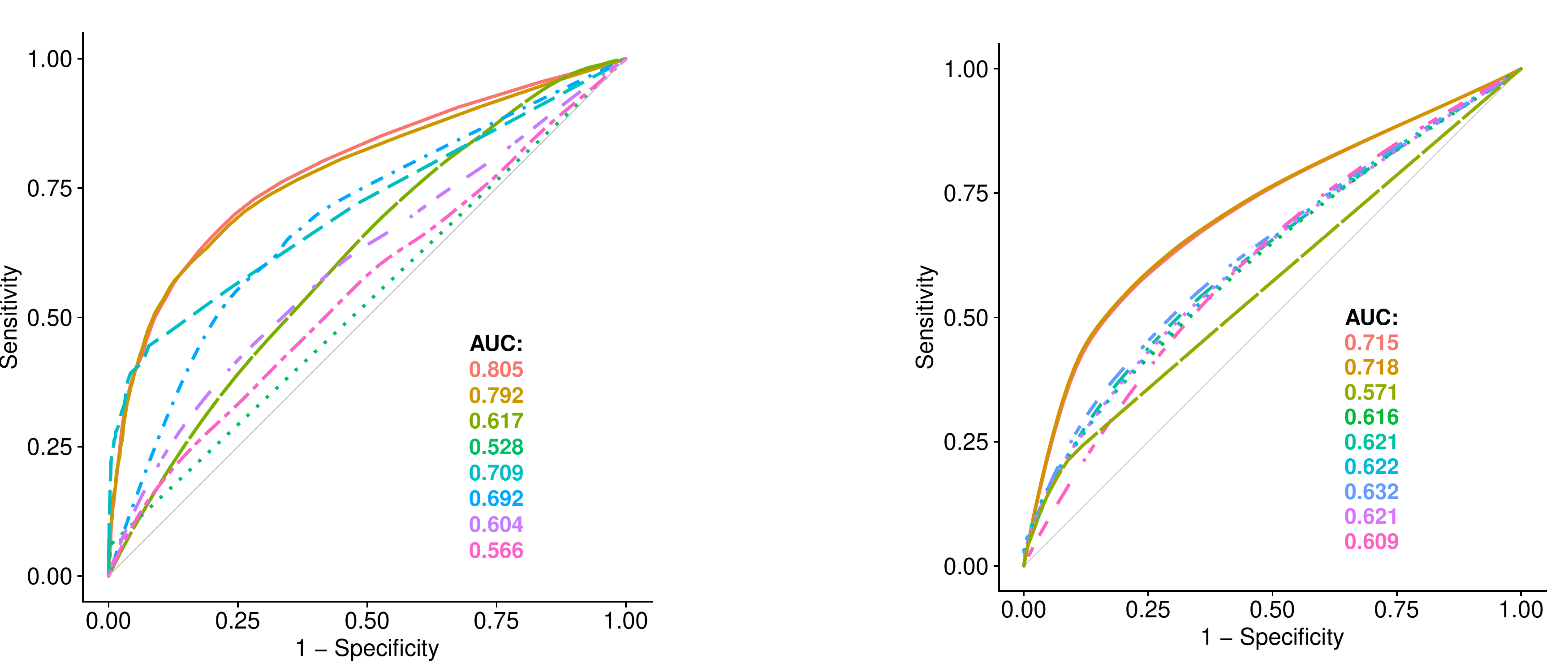}\label{size20}}
	\end{tabular}
	\begin{tabular}{@{}cc@{}}
		\centering
		\subfloat{\includegraphics[width=0.55\textwidth]{lg_gamma.pdf}} & 
		\subfloat{\includegraphics[width=0.5\textwidth]{lg_delta.pdf}}
	\end{tabular}
	\caption[]{\bch Averaged ROC curves for the discriminating feature indicator $\bm{\gamma}$ (left) and the feature-covariate association indicator $\bm{\Delta}$ (right) with respect to different sample sizes per group (a) $n / 2 = 30$ and (b) $10$, over $100$ replicates in each scenario.\ech  }
	\label{sample_supp}
\end{figure}

\begin{figure}[!h]
	\centering
	\begin{tabular}{@{}c@{}}
		\subfloat[][ROC curves for $\bm{\gamma}$ (left) and $\bm{\Delta}$ (right) with a log-scale noise level of $\sigma_e = 0.5$]{\includegraphics[width=0.9\textwidth]{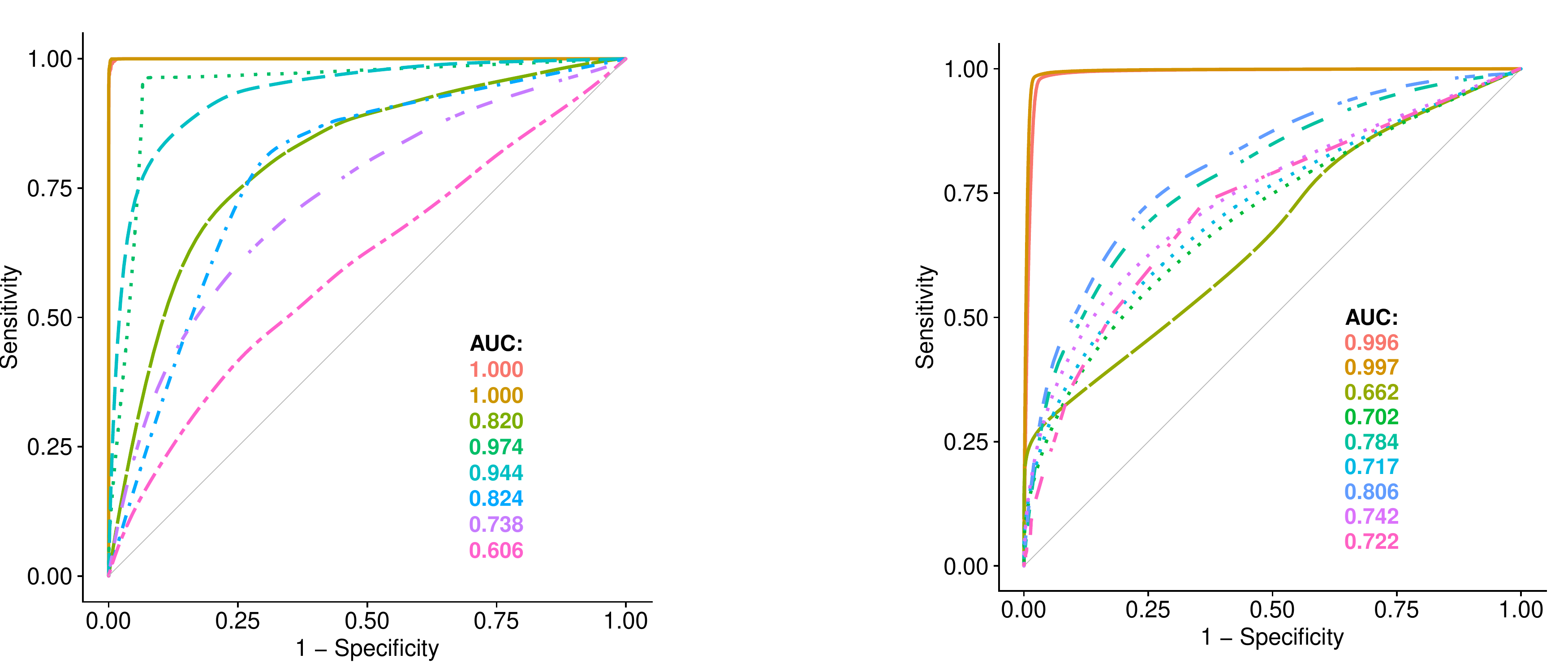}} \\
		\subfloat[][ROC curves for $\bm{\gamma}$ (left) and $\bm{\Delta}$ (right) with a log-scale noise level of $\sigma_e = 1.0$]{\includegraphics[width=0.9\textwidth]{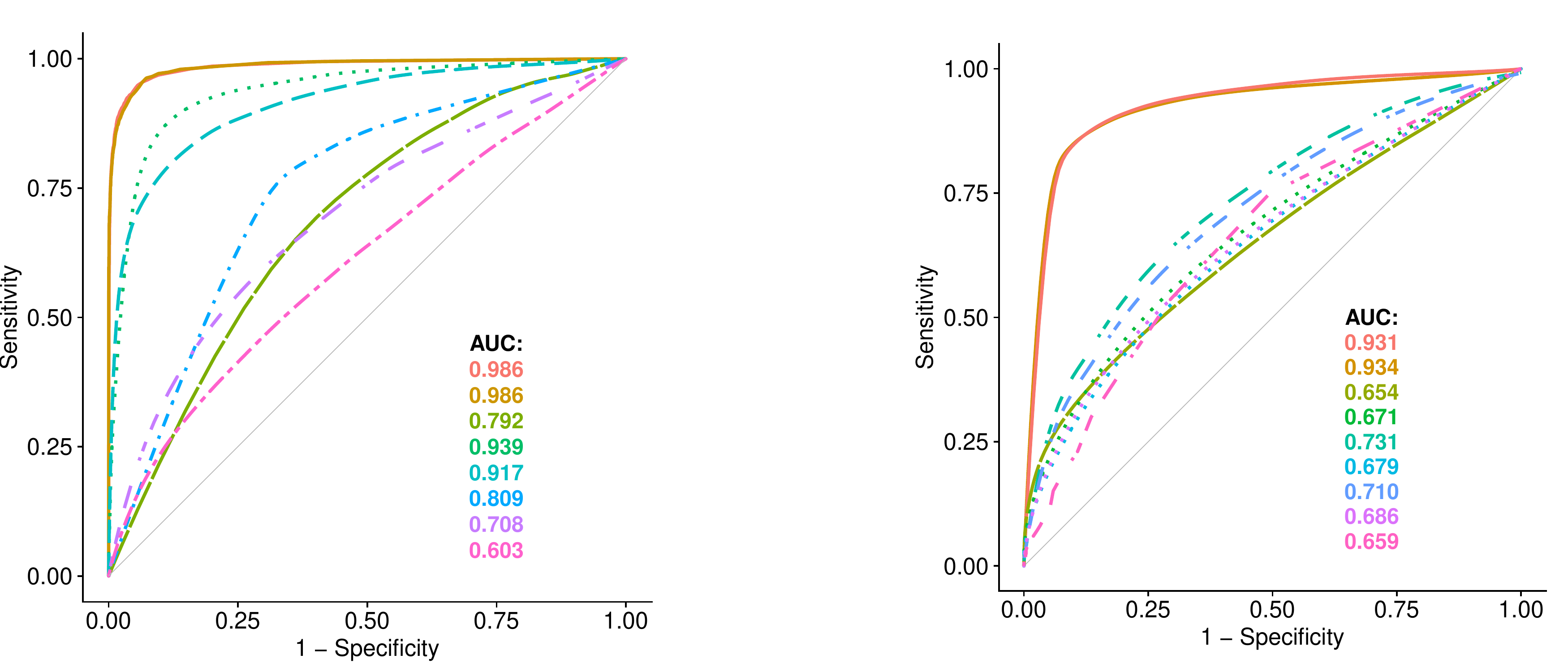}}\\
		\subfloat[][ROC curves for $\bm{\gamma}$ (left) and $\bm{\Delta}$ (right) with a log-scale noise level of $\sigma_e = 1.5$]{\includegraphics[width=0.9\textwidth]{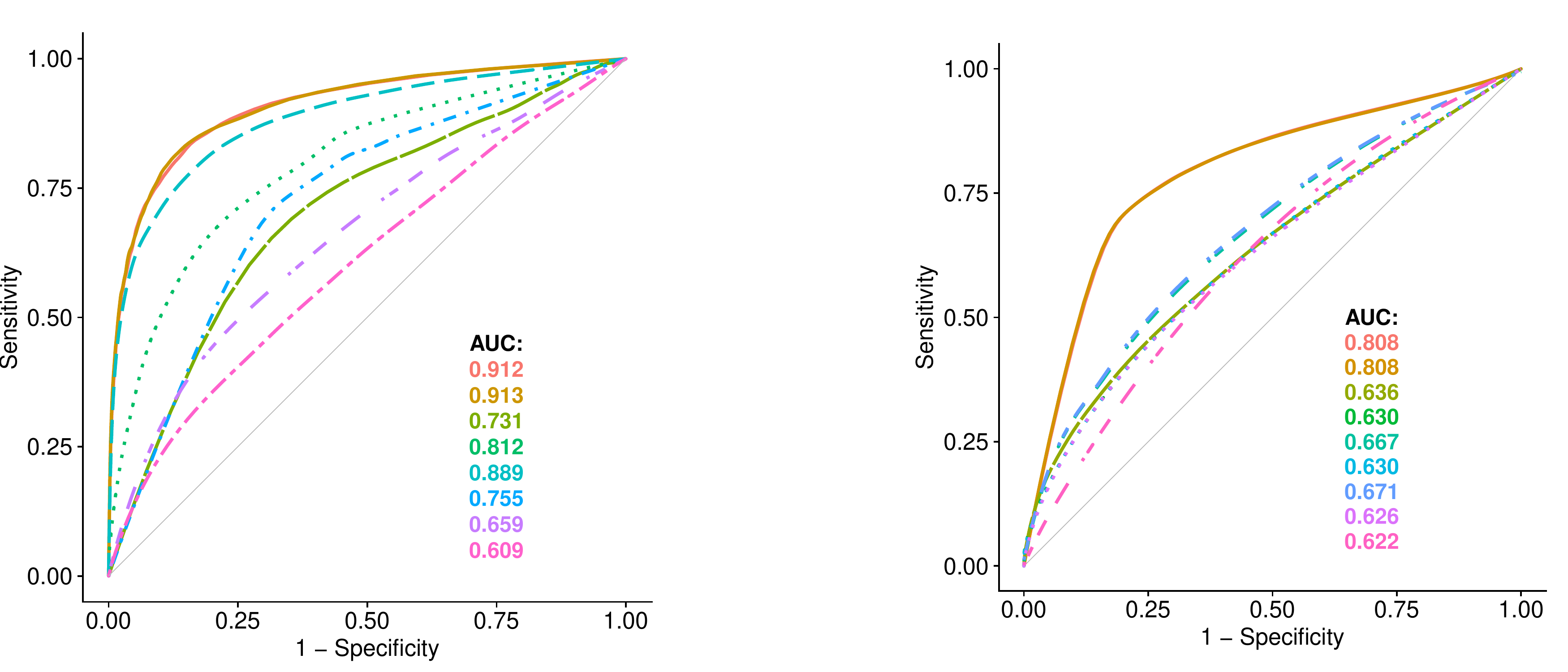}}
	\end{tabular}
	\begin{tabular}{@{}cc@{}}
		\centering
		\subfloat{\includegraphics[width=0.55\textwidth]{lg_gamma.pdf}} & 
		\subfloat{\includegraphics[width=0.5\textwidth]{lg_delta.pdf}}
	\end{tabular}
	\caption[]{\bch Averaged ROC curves for the discriminating feature indicator $\bm{\gamma}$ (left) and the feature-covariate association indicator $\bm{\Delta}$ (right) with respect to different noise levels (a) $\sigma_e=0.5$, (b) $\sigma_e= 1.0$, and (c) $\sigma_e=1.5$, over $100$ replicates in each scenario. \ech }
	\label{noise_supp}
\end{figure}

\begin{figure}[!h]
	\centering
	\begin{tabular}{@{}c@{}}
		\subfloat[][ROC curves for $\bm{\gamma}$ (left) and $\bm{\Delta}$ (right) with a false zero proportion of $\pi_0 = 30\%$]{\includegraphics[width=0.9\textwidth]{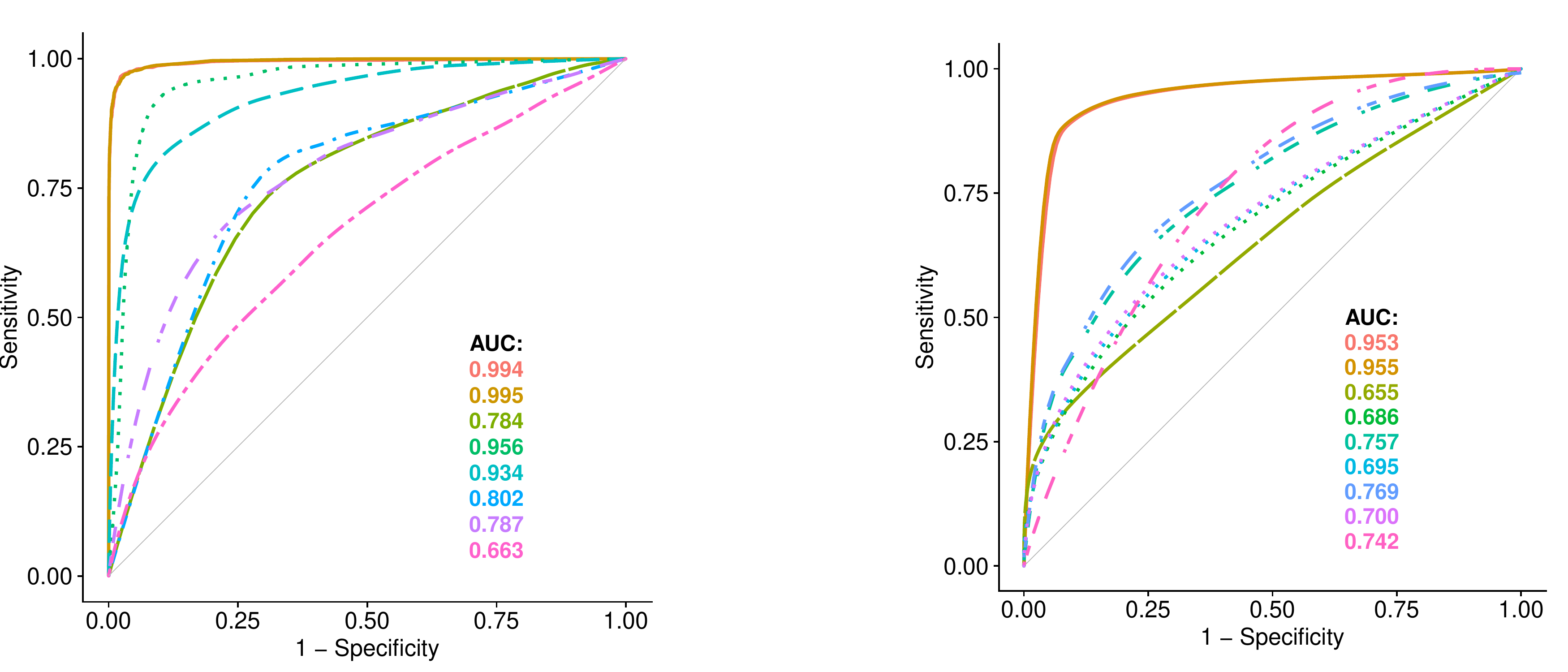}} \\
		\subfloat[][ROC curves for $\bm{\gamma}$ (left) and $\bm{\Delta}$ (right) with a false zero proportion of $\pi_0 = 40\%$]{\includegraphics[width=0.9\textwidth]{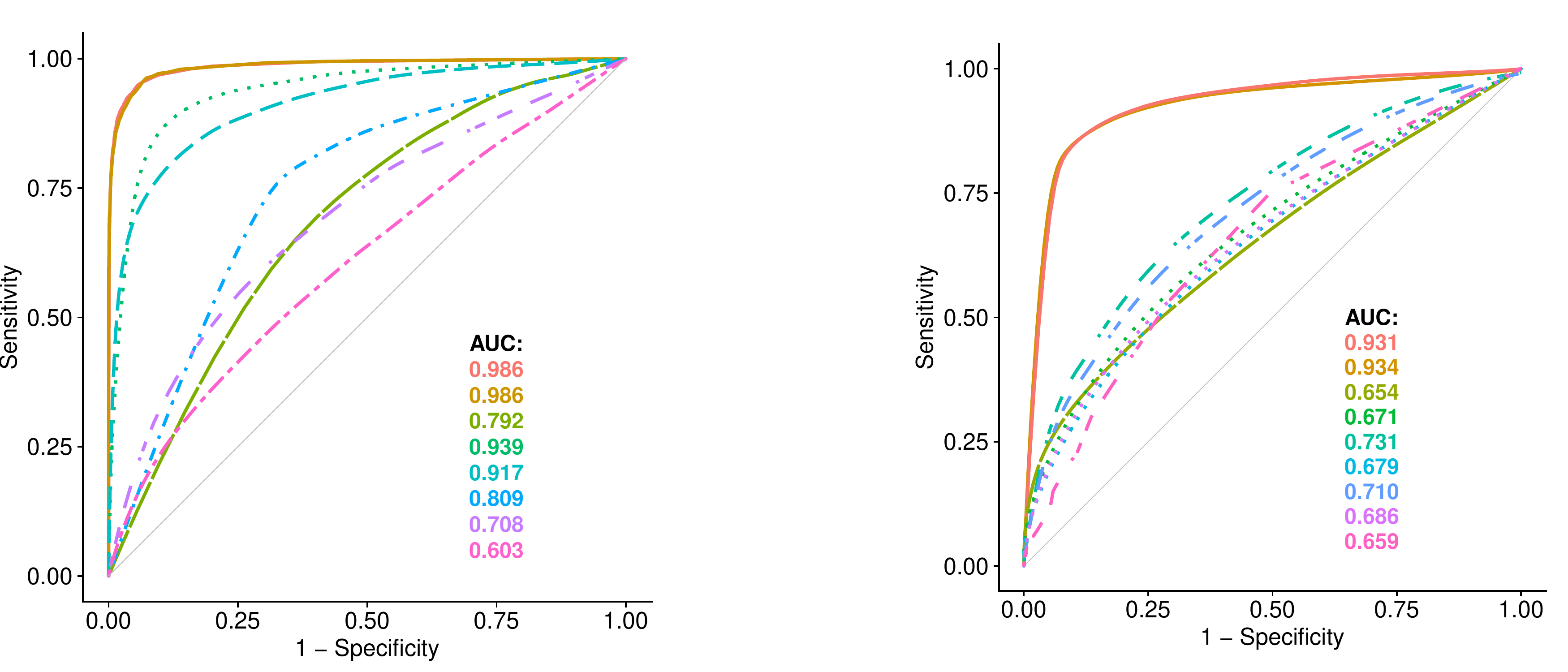}}\\
		\subfloat[][ROC curves for $\bm{\gamma}$ (left) and $\bm{\Delta}$ (right) with a false zero proportion of $\pi_0 = 70\%$]{\includegraphics[width=0.9\textwidth]{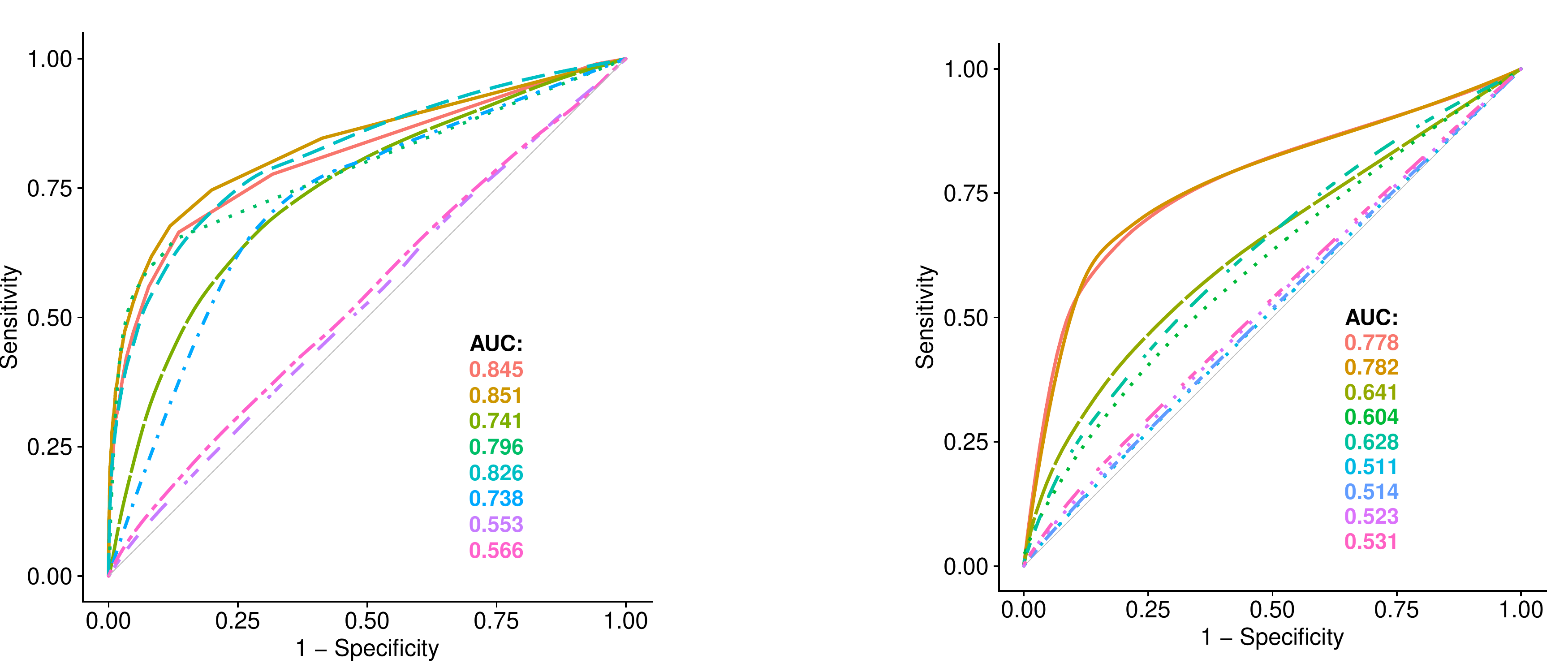}}
	\end{tabular}
	\begin{tabular}{@{}cc@{}}
		\centering
		\subfloat{\includegraphics[width=0.55\textwidth]{lg_gamma.pdf}} & 
		\subfloat{\includegraphics[width=0.5\textwidth]{lg_delta.pdf}}
	\end{tabular}
	\caption[]{\bch Averaged ROC curves for the discriminating feature indicator $\bm{\gamma}$ (left) and the feature-covariate association indicator $\bm{\Delta}$ (right) with respect to different false zero proportions (a) $\pi_0=30\%$, (b) $\pi_0=40\%$, and (c) $\pi_0=70\%$, over $100$ replicates in each scenario.\ech }
	\label{zero_supp}
\end{figure}

\newpage
\begin{table}[!h]
	\centering
	\hspace*{-1cm}
	\scalebox{0.7}{
		\renewcommand{\arraystretch}{1.2}
		\begin{tabular}{lcccccccc}
			\hline  
			\multicolumn{1}{c}{Liver Cirrhosis}& \multicolumn{2}{c}{Major Parameter }& \multicolumn{2}{c}{Covariate Effect} & \multicolumn{2}{c}{Estimation for} & \multicolumn{2}{c}{Estimation for}   \\ 
			\multicolumn{1}{c}{Study} & \multicolumn{2}{c}{Estimation }& \multicolumn{2}{c}{by Group } & \multicolumn{2}{c}{Health} & \multicolumn{2}{c}{Liver Cirrhosis}   \Tstrut \\ 
			\hline
			\multicolumn{1}{c}{Discriminatory Taxa} & $\mu_{0j}$ & $\mu_{2j}$ & $\text{Ave}(\bm{x}\boldsymbol{\beta}^T)$ & $\text{Ave}(\bm{x}\boldsymbol{\beta}^T)$ & Normalized & Estimated & Normalized & Estimated \\ 
			\multicolumn{1}{c}{(ordered by posterior mean of $\mu_{2j}$)} & (CI for $\mu_{0j}$) & (CI for $\mu_{2j}$)& Health & Liver Cirrhosis & $\log(y_{\cdot j})$  & $\alpha_{\cdot j}$ &  $\log(y_{\cdot j})$ & $\alpha_{\cdot j}$ \Tstrut \\ 
			\hline  
			\textit{Bacteroides eggerthii} & 12.93 & -0.23 & 0.47 & -0.37 & 10.79 & 13.40 & 8.82 & 12.32 \\
			& (12.77, 13.11) & (-0.38, -0.08) &   &   &   &   &  &   \\  
			\textit{Gammaproteobacteria} & 14.49 & 0.22 & -0.54 & 0.50 & 12.52 & 13.95 & 14.24 & 15.22 \\ 
			& (14.31, 14.70) & (0.07, 0.37) &   &   &   &   &  &   \\ 
			\textit{Veillonella dispar} & 11.18 & 0.22 & -0.84 & 0.75 & 8.14 & 10.34 & 10.79 & 12.16 \\ 
			& (10.96, 11.42) & (0.06, 0.38) &   &   &   &   &  &   \\ 
			\textit{Bacteroides caccae} & 13.37 & 0.23 & -0.11 & 0.12 & 11.48 & 13.26 & 11.88 & 13.72 \\ 
			& (13.19, 13.56) & (0.08, 0.38)&   &   &   &   &  &   \\ 
			\textit{Streptococcus salivarius} & 12.28 & 0.26 & -0.67 & 0.63 & 10.22 & 11.61 & 12.25 & 13.17 \\ 
			&(12.10, 12.47)& (0.11, 0.41) &   &   &   &   &  &   \\ 
			\textit{Haemophilus parainfluenzae} & 12.98 & 0.26 & -0.57 & 0.52 & 10.12 & 12.40 & 12.51 & 13.76 \\ 
			& (12.80, 13.18) &(0.12, 0.40)&   &   &   &   &  &   \\ 
			\textit{Haemophilus} & 13.00 & 0.27 & -0.57 & 0.52 & 10.13 & 12.43 & 12.53 & 13.79 \\ 
			&(12.83, 13.19)& (0.12, 0.41) &   &   &   &   &  &   \\ 
			\textit{Streptococcus parasanguinis} & 11.48 & 0.28 & -0.82 & 0.76 & 8.70 & 10.66 & 11.38 & 12.53 \\
			&(11.29, 11.70) & (0.12, 0.42) &   &   &   &   &  &   \\  
			\textit{Pasteurellales} & 13.04 & 0.29 & -0.57 & 0.52 & 10.14 & 12.47 & 12.56 & 13.84 \\ 
			&(12.86, 13.22)& (0.13, 0.42) &   &   &   &   &  &   \\ 
			\textit{Pasteurellaceae} & 13.03 & 0.29 & -0.57 & 0.52 & 10.14 & 12.47 & 12.56 & 13.84 \\ 
			& (12.86, 13.2) & (0.13, 0.43) &   &   &   &   &  &   \\ 
			\textit{Veillonella parvula} & 12.01 & 0.29 & -0.63 & 0.63 & 9.39 & 11.38 & 11.96 & 12.93 \\ 
			& (11.83, 12.21) &(0.13, 0.43) &   &   &   &   &  &   \\ 
			\textit{Streptococcus} & 12.88 & 0.32 & -0.66 & 0.61 & 10.61 & 12.22 & 13.17 & 13.81 \\ 
			& (12.71, 13.05)& (0.17, 0.45) &   &   &   &   &  &   \\ 
			\textit{Streptococcaceae} & 12.89 & 0.32 & -0.66 & 0.61 & 10.61 & 12.22 & 13.19 & 13.82 \\ 
			& (12.72, 13.07) & (0.17, 0.46) &   &   &   &   &  &   \\ 
			\textit{Klebsiella pneumoniae} & 13.09 & 0.32 & -0.39 & 0.42 & 10.38 & 12.71 & 11.78 & 13.84 \\ 
			& (12.91, 13.28)& (0.17, 0.46) &   &   &   &   &  &   \\ 
			\textit{Bacilli} & 13.36 & 0.34 & -0.71 & 0.66 & 11.03 & 12.64 & 13.59 & 14.36 \\ 
			& (13.19, 13.53) & (0.19, 0.47) &   &   &   &   &  &   \\ 
			\textit{Klebsiella} & 13.19 & 0.34 & -0.45 & 0.48 & 10.32 & 12.74 & 11.69 & 14.01 \\ 
			& (13.00, 13.37) & (0.20, 0.48) &   &   &   &   &  &   \\ 
			\textit{Lactobacillales} & 13.33 & 0.35 & -0.71 & 0.66 & 10.88 & 12.62 & 13.58 & 14.34 \\ 
			&(13.16, 13.51) &(0.20, 0.49) &   &   &   &   &  &   \\ 
			\textit{Veillonella} & 13.47 & 0.39 & -0.81 & 0.75 & 10.27 & 12.66 & 13.88 & 14.61 \\ 
			& (13.30, 13.65) & (0.25, 0.52) &   &   &   &   &  &   \\ 
			\textit{Veillonella unclassified} & 12.93 & 0.40 & -0.88 & 0.83 & 9.28 & 12.05 & 13.27 & 14.17 \\ 
			&(12.75, 13.13) & (0.26, 0.55) &   &   &   &   &  &   \\ 
			\hline
		\end{tabular}
	}
	\caption{Liver cirrhosis dataset: parameter estimation for the identified discriminating taxa from the liver cirrhosis study. Posterior mean and 95\% Credible Interval (CI) are reported for the estimated $\mu_{0j}$ (feature-specific baseline parameter) and $\mu_{2j}$(group-specific parameter); Covariate effect represents the mean of $\bm{x}\hat{\boldsymbol{\beta}^T}$ of all samples in the corresponding patient group; Normalized $\log(y_{\cdot j})$ is the mean of log scaled observations after accounting for the sample heterogeneity factor (i.e. size factor) $s_i$. Estimated $\alpha_{\cdot j}$ is the mean of $\alpha_{ij}$ for all sample $i$ from the same patient group.}
	\label{table1}
\end{table}

\begin{table}[ht]
	\centering
	\hspace*{-1cm}
	\scalebox{0.7}{
		\renewcommand{\arraystretch}{1.2}
		\begin{tabular}{lcccccccc}
			\hline  
			\multicolumn{1}{c}{Liver Cirrhosis}& \multicolumn{2}{c}{Major Parameter }& \multicolumn{2}{c}{Covariate Effect} & \multicolumn{2}{c}{Estimation for} & \multicolumn{2}{c}{Estimation for}   \\ 
			\multicolumn{1}{c}{Study} & \multicolumn{2}{c}{Estimation }& \multicolumn{2}{c}{by Group } & \multicolumn{2}{c}{Health} & \multicolumn{2}{c}{Liver Cirrhosis}   \Tstrut \\ 
			\hline
			\multicolumn{1}{c}{Discriminatory Taxa} & $\mu_{0j}$ & $\mu_{2j}$ & $\text{Ave}(\bm{x}\boldsymbol{\beta}^T)$ & $\text{Ave}(\bm{x}\boldsymbol{\beta}^T)$ & Normalized & Estimated & Normalized & Estimated \\ 
			\multicolumn{1}{c}{(ordered by posterior mean of $\mu_{2j}$)} & (CI for $\mu_{0j}$) & (CI for $\mu_{2j}$)& Health & Liver Cirrhosis & $\log(y_{\cdot j})$  & $\alpha_{\cdot j}$ &  $\log(y_{\cdot j})$ & $\alpha_{\cdot j}$ \Tstrut \\ 
			\hline  
			\textit{Bifidobacterium} & 12.50 & -1.34 & -0.19 & 0.62 & 10.78 & 12.31 & 10.97 & 11.79  \\ 
			& (12.16, 12.86) & (-2.08, -0.62) &   &   &   &   &  &   \\ 
			\textit{Bifidobacteriaceae} & 12.50 & -1.34 & -0.19 & 0.62 & 10.78 & 12.31 & 10.97 & 11.78 \\ 
			&(12.16, 12.86) &(-2.07, -0.62)&   &   &   &   &  &   \\ 
			\textit{Bifidobacteriales} & 12.50 & -1.33 & -0.19 & 0.61 & 10.78 & 12.31 & 10.97 & 11.78 \\ 
			& (12.16, 12.86) & (-2.06, -0.60) &   &   &   &   &  &   \\ 
			\textit{Clostridium~methylpentosum} & 8.13 & 0.94 & 0.18 & -0.39 & 8.21 & 8.31 & 8.69 & 8.68 \\ 
			&(5.95, 6.90) & (0.54, 1.34) &   &   &   &   &  &   \\ 
			\textit{Carnobacteriaceae}& 6.40 & 1.03 & 0.31 & -0.24 & 6.56 & 6.71 & 7.12 & 7.18 \\ 
			&(10.13, 11.22) & (0.56, 1.45) &   &   &   &   &  &   \\ 
			\textit{Clostridium~bartlettii} & 10.60 & 1.40 & -0.28 & 0.46 & 9.35 & 10.32 & 11.19 & 12.46 \\  
			& (7.88, 8.54)&(0.65, 2.15) &   &   &   &   &  &   \\ 
			\textit{Eubacterium~siraeum} & 11.84 & 1.42 & 0.30 & -0.35 & 10.70 & 12.14 & 12.41 & 12.92 \\ 
			& (11.35, 12.42)&(0.64, 2.18) &   &   &   &   &  &  \\
			\hline
		\end{tabular}
	}
	\caption{Metastatic melanoma dataset: parameter estimation for the identified discriminating taxa from the metastatic melanoma study. Posterior mean and 95\% Credible Interval (CI) are reported for the estimated $\mu_{0j}$ (feature-specific baseline parameter) and $\mu_{2j}$(group-specific parameter); Covariate effect represents the mean of $\bm{x}\hat{\boldsymbol{\beta}^T}$ of all samples in the corresponding patient group; Normalized $\log(y_{\cdot j})$ is the mean of log scaled observations after accounting for the sample heterogeneity factor (i.e. size factor) $s_i$. Estimated $\alpha_{\cdot j}$ is the mean of $\alpha_{ij}$ for all sample $i$ from the same patient group.}
	\label{table2}
\end{table}

\clearpage

\begin{figure}[!h]
	\centering
	\includegraphics[width = \linewidth]{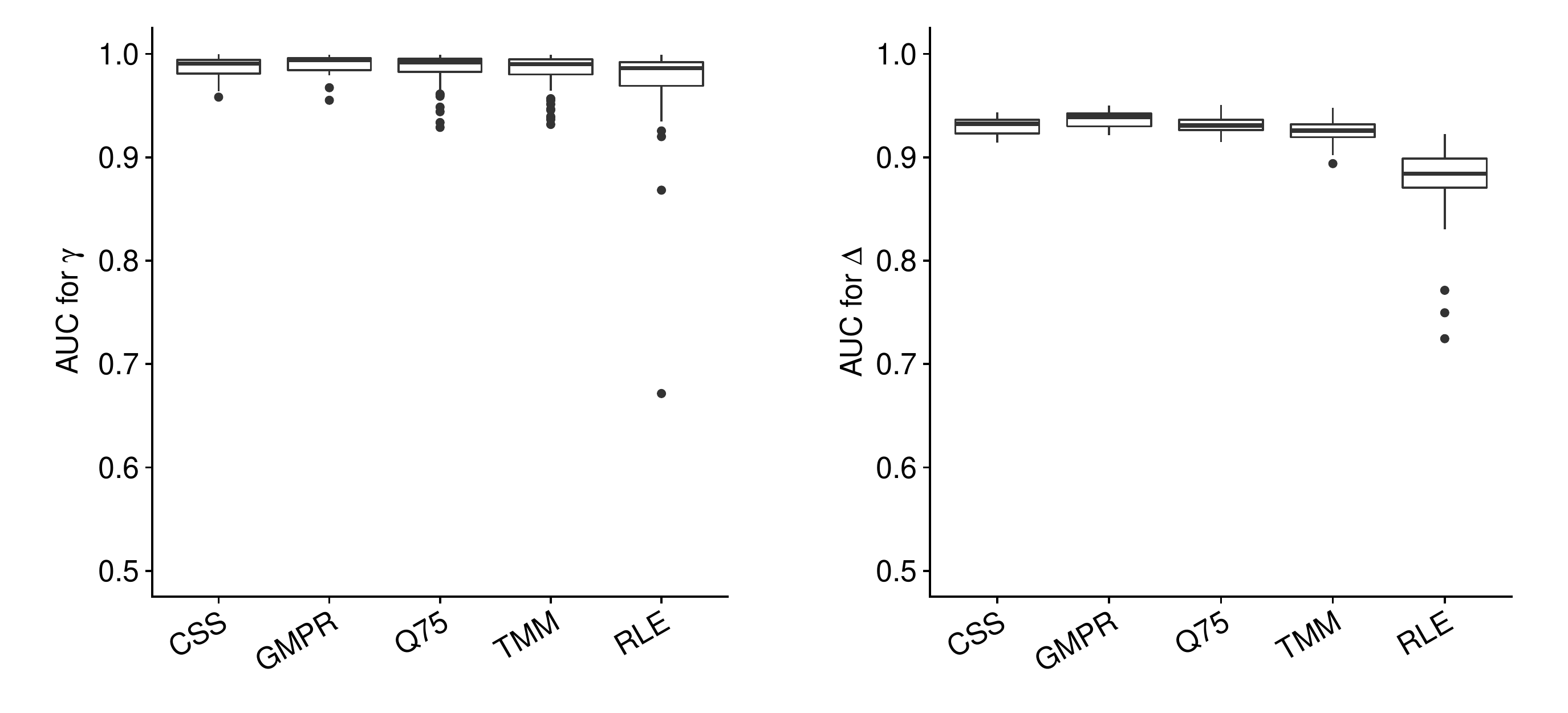}
	\caption{\bch Side-by-side box plots of AUCs for the discriminating feature indicator $\bm{\gamma}$ (left) and the feature-covariate association indicator $\bm{\Delta}$ (right) with respect to different normalization techniques, over $100$ reference simulated datasets. CSS for cumulative sum scaling. GMPR for geometric mean of pairwise ratios. Q75 for 0.75-th quantile. TMM for trimmed mean of M values. RLE for relative log expression. \ech }
	\label{size_factor_fig}
\end{figure}

\begin{figure}[!h]
	\centering
	\includegraphics[width = 0.6\linewidth]{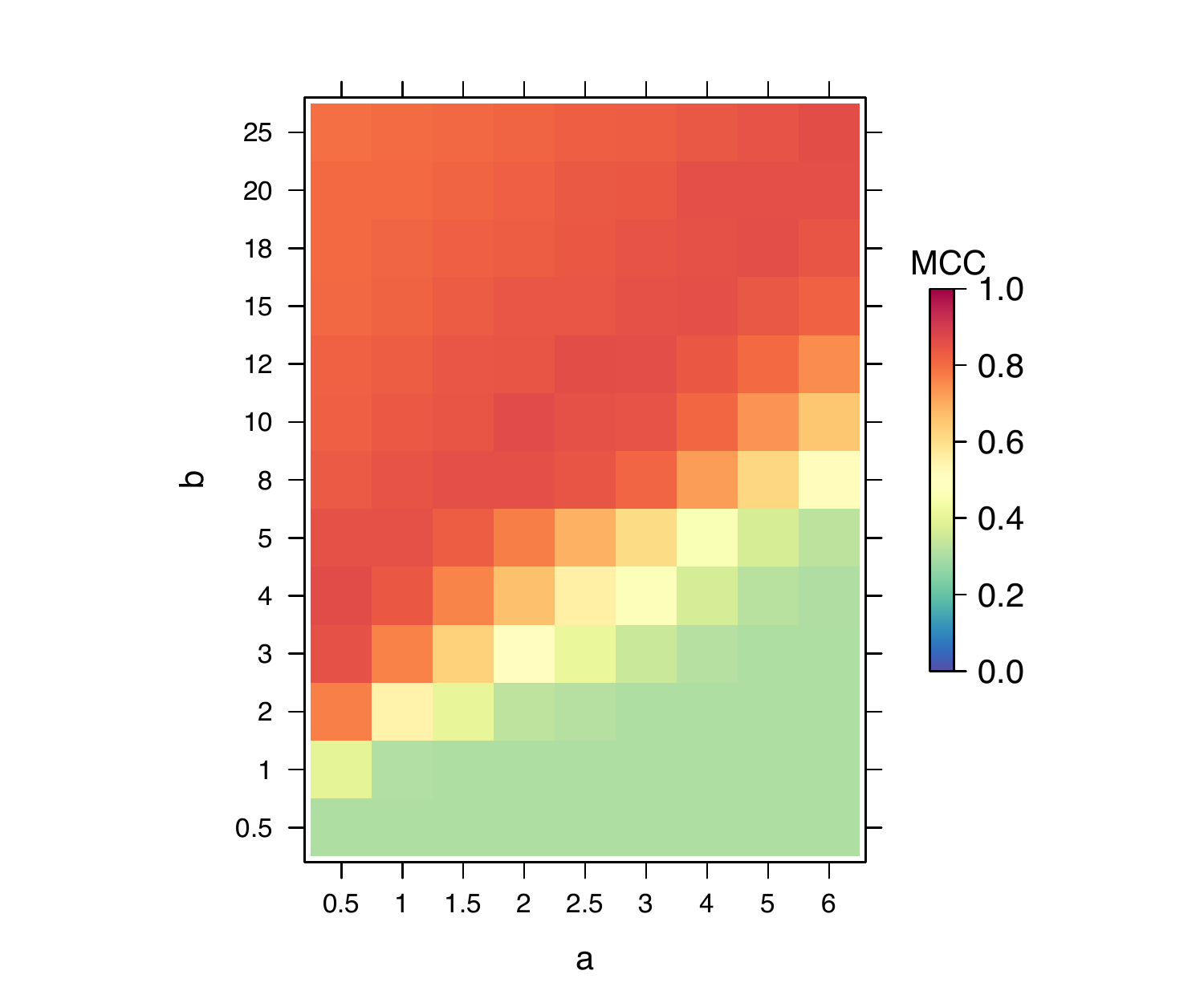}
	\caption{Heatmap of Matthews correlation coefficients (MCC) for the discriminating feature indicator $\bm{\gamma}$ with the choice of ($a, b$) from the inverse-gamma prior on the variance terms $\sigma_{\mu  j}^2$ and $\sigma_{\beta j}^2$. Each value of MCC represents the averaged result of 30 replicates. }
	\label{sensi}
\end{figure}
\end{document}